\documentclass[preprint,12pt]{elsarticle} 
\usepackage{booktabs}
\usepackage{amssymb}
\usepackage[colorlinks]{hyperref}   
\usepackage{xcolor}
\usepackage{amsthm}
\usepackage{mathtools}
\usepackage{enumitem}
\usepackage{graphicx}
\usepackage{subfigure}
\usepackage{autobreak}
\usepackage{lineno}
\usepackage{natbib}
\usepackage[margin=2cm]{geometry}
\usepackage{multirow}
\usepackage{floatrow}
\usepackage{colortbl}
\usepackage{threeparttable}
\usepackage{subcaption}

\usepackage{graphicx}
\newfloatcommand{capbtabbox}{table}[][\FBwidth]
\usepackage{multirow}
\usepackage{amsthm,amsmath,amssymb}
\usepackage{mathrsfs}
\usepackage{bm}

\usepackage[ruled,linesnumbered]{algorithm2e}

\setcitestyle{authoryear,open={(},close={)}}
\definecolor{orange}{rgb}{1,0.5,0}
\definecolor{red}{RGB}{198,0,35}
\definecolor{amberseldef}{rgb}{1.0, 0.49, 0.0}
\definecolor{ceruleanblue}{rgb}{0.16, 0.32, 0.75}
\definecolor{amber}{rgb}{1.0, 0.49, 0.0}
\definecolor{dodgerblue}{rgb}{0.12, 0.56, 1.0}
\definecolor{pureblue}{rgb}{0, 0, 1.0}

\definecolor{blue}{rgb}{0.0, 0.28, 0.67}

\def\hmath$#1${\texorpdfstring{{\rmfamily\textit{#1}}}{#1}}

\makeatletter\href{}{}
\def\ps@pprintTitle{
   \let\@oddhead\@empty
   \let\@evenhead\@empty
   \let\@oddfoot\@empty
   \let\@evenfoot\@oddfoot
}
\makeatother

\AtBeginDocument{\hypersetup{citecolor= pureblue,linkcolor = pureblue,urlcolor = dodgerblue}}
\begin{document}


\begin{frontmatter}


\address[1]{Department of Civil Engineering, The University of Hong Kong, Hong Kong, China}
\address[2]{Department of Urban Planning and Design, The University of Hong Kong, Hong Kong, China}
\cortext[cor1]{Corresponding author: guanhuang@connect.hku.hk}

\author[1]{Wang Chen}
\author[2]{Guan Huang \texorpdfstring{\corref{cor1}}{}}
\author[1]{Jintao Ke}

\title{Development dilemma of ride-sharing: Revenue or social welfare?}

\begin{abstract}
This study investigates the development dilemma of ride-sharing services using real-world mobility datasets from nine cities and calibrated customers' price and detour elasticity. Through massive numerical experiments, this study reveals that while ride-sharing can benefit social welfare, it may also lead to a loss of revenue for transportation network companies (TNCs) or drivers compared with solo-hailing, which limits TNCs' motivation to develop ride-sharing services. Three key factors contributing to this revenue loss are identified: (1) the low successful sharing ratio for customers choosing ride-sharing in some cases, (2) the limited saved trip distance by pooling two customers, and (3) the potential revenue loss when pooling customers with significantly different trip fares. Furthermore, this study finds that the monetary benefits of carbon emission reductions from ride-sharing are not substantial enough to affect customers' choices between solo-hailing and ride-sharing. The findings provide a valuable reference for TNCs and governments. For TNCs, effective pricing strategies, such as dynamic pricing, should be designed to prevent revenue loss when introducing ride-sharing. Governments are suggested to subsidize ride-sharing services to solve the development dilemma and maintain or even increase social welfare benefits from ride-sharing, including reduced carbon emissions and improved vehicle occupancy rates.

\end{abstract}

\begin{keyword}
Ride-sharing; revenue; social welfare; development dilemma; carbon emissions
\end{keyword}

\end{frontmatter}

\section{Introduction}

Ride-sharing is a promising mobility service that matches multiple customers in a similar direction and serves them in one vehicle. Due to the shared usage of vehicles, it increases vehicle occupancy and transport efficiency. On the one hand, ride-sharing is believed to let the vehicle operator serve more customers simultaneously and earn more revenues \citep{fagnant2018dynamic}. On the other hand, ride-sharing can create more social welfare by providing customers a split fare and shorter total travel time \citep{shaheen2020going, fagnant2018dynamic}, reducing traffic congestion and traffic emissions with a smaller fleet size \citep{li2021does, chen2024quantifying}. Currently, many transportation network companies (TNCs) provide ride-sharing services, such as Uber Pool, Lyft Shared, DiDi Express carpool, and GrabShare.

Although ride-sharing is deemed to be a win-win mobility service in creating more revenues for operators and more social welfare for customers and society, the current situation of ride-sharing is that it is only a niche in the mobility service market. For example, ride-sharing orders only account for 5.5\% of total orders in Chengdu, China \citep{huang2021spatiotemporally}, and can rise to 15\% in Toronto \citep{young2020true}. However, even in the most well-developed shared mobility market such as Manhattan, New York City, the penetration is less than a quarter \citep{huang2024shareable}. 

Ride-sharing services have significant economies of scale where service performance, e.g. shareability, cost, and detour are found to benefit from the increase in ridership \citep{huang2024shareable, ke2021data, liu2023scale, chen2023scaling}. Therefore, it is theoretically possible to achieve a virtuous circle in the ride-sharing services that ``higher adoption'' $\rightarrow$ ``better service performance'' $\rightarrow$ ``higher adoption''. However, the aforementioned low adoption rate proves that the circle is not effectively running, indicating that ride-sharing is under a ``rich in expectation but poor in reality'' status quo.

Improving the adoption of ride-sharing is a key to breaking the development dilemma. In terms of the determinants of ride-sharing adoption, extensive prior studies proved that the discomfort of sharing vehicle space with strangers, and particularly the trade-off between the discount in fare and the delay due to picking up/dropping off the trip partner, are the two main influencing factors \citep{huang2023share, alonso2021determinants, lavieri2019modeling}. However, TNCs, as profit-seeking organizations, cannot offer excessively low discounts to attract customers. Therefore, it is of great importance to investigate the relationship between TNCs' fare discounts and their service performances and revenue. In this study, we stand at the operator's perspective, proposing a comprehensive simulation framework, including pricing, matching, and repositioning modules to explore how TNCs' ride-sharing operation strategies will influence their revenue and social welfare indicators, such as customers' waiting times, vehicle occupancy rates, and carbon emissions. The simulations are conducted on an agent-based simulation platform with real-world mobility datasets from nine cities in China. In addition, a survey is incorporated to calibrate the elasticity of customers towards price and detour. Differing from the prior studies \citep{alonso2017demand, lokhandwala2018dynamic}, this study stands at the TNCs' perspective because, as the ride-sharing service operator, it is TNCs' revenue that determines whether they have the motivation to develop this service. The findings of this study can answer whether ride-sharing can truly improve revenue and enhance social welfare, providing TNCs with strategy implications to effectively drive the positive circle of ride-sharing. The main contributions of this study are summarized as follows:

\begin{itemize}
    \item This study proposes an agent-based simulation framework, including pricing, matching, and repositioning modules, which can simulate solo-hailing and ride-sharing services simultaneously on large-scale road networks.
    
    \item Through extensive experiments based on real-world mobility datasets and customers' price and detour elasticity, this study discovers the development dilemma of ride-sharing: compared with the pure solo-hailing service, introducing ride-sharing can enhance social welfare but can also induce a loss of revenue.
    
    \item This study reveals three key factors that can result in revenue loss. First, the platform successfully pools customers who choose ride-sharing at a relatively low rate, especially in areas with low demand. Second, compared with accommodating two customers separately, the average saved trip distance by pooling the two customers is limited. Third, it is not revenue-effective for the platform to pool two customers with large price differences, as the platform can only serve the customer with the higher price without giving a discount in a pure solo-hailing system.
    
    \item This study thoroughly discusses the potential compensation methods for revenue loss when introducing ride-sharing. The results indicate that the saved carbon emissions from ride-sharing cannot compensate for its revenue loss. Instead, TNCs should design efficient pricing algorithms for ride-sharing and solo-hailing services to improve their revenue. Also, government subsidies can benefit ride-sharing services, enhancing the social welfare. This study provides valuable insights into designing effective strategies and policies for ride-sharing services.
\end{itemize}

The remainder of this paper is organized as follows. The studies related to ride-sharing quantification and simulations are first reviewed in Section \ref{sec: LR}. Subsequently, the methodology used in this study is explicitly introduced in Section \ref{sec: method}, and the experiments and results are presented and discussed in Section \ref{sec: exp}. Then, a typical scenario is used to further analyze ride-sharing services, and a potential solution to prevent the revenue loss is discussed in Section \ref{sec: discussion}. Finally, the main conclusions of this study and future research directions are summarized in Section \ref{sec: conclusion}.


\section{Literature review}\label{sec: LR}

\subsection{Ride-sharing and its development}
Ride-sharing services match multiple customers in a similar direction and serve them in one vehicle. With the shared usage of vehicles, it is expected to be a promising future mobility revolution to reduce the urban transportation system vehicle fleet size and alleviate traffic congestion and emissions \citep{hyland2020operational, chen2024quantifying}. 

Ride-sharing is not an entirely newly-emerged concept. The earliest vehicle-sharing concept was proposed in the last century, presented as a car-sharing club and carpooling program \citep{chan2012ridesharing}. Prior studies explored the reasons behind the rise and fall of carpooling services from several aspects and provided evidence about its development. Policy support is an important driving force. Historically, the emergence of car sharing is due to the promotion of the US government in World War II to conserve gasoline and rubber \citep{chan2012ridesharing}. The most essential factor is the economy. For instance, the high gasoline price significantly contributed to the growth of carpooling programs in the 1970s \citep{pratsch1975carpool}. Apart from the energy price, the non-monetary time cost also matters. For example, the introduction of the high-occupancy-vehicle (HOV) lane, which ensures a higher traffic speed for the shared vehicle increased the attraction of carpooling \citep{chang2008review}, while urban expansion in the last century, which led to a greater sharing detour decreased the adoption of this service \citep{ferguson1970rise}.

In recent years, the development of communication technologies shared mobility services, and artificial intelligence led to the renascence and development of ride-sharing. Through convenient app-based mobility services and advanced algorithms, TNCs can efficiently provide optimal ride-sharing routes which solve some of the prior barriers \citep{narayanan2020shared}. Besides, the potential of ride-sharing to alleviate severe urban traffic problems also attracts policymakers' attention. However, although the potential benefits of ride-sharing are repeatedly emphasized \citep{alonso2017demand, lokhandwala2018dynamic, hyland2020operational, fagnant2014travel}, it still has limited adoption after years of development. Although prior studies provided implications from the customer's choice perspective \citep{huang2023share, alonso2021determinants, lavieri2019modeling}, seldom focused on the interests of TNCs. Therefore, it is important to understand the dilemma and barriers of ride-sharing development from the TNCs' perspective for a more comprehensive solution.

\subsection{Quantitative study on ride-sharing benefits and performance}

Due to the limited adoption and deployment of the ride-sharing service, and access to its operation data, the quantitative study on ride-sharing benefits and performance is mainly based on equilibrium theory and simulation modeling. Equilibrium theory-based analysis is widely used in transportation studies by considering a series of exogenous and endogenous variables. In terms of ride-sharing, \cite{xu2015complementarity} proposed a traffic assignment model that incorporates ride-sharing as a transportation mode to analyze the relationship between user acceptance, traffic congestion, and ride-sharing under the framework of the Wardrop traffic equilibrium model. \cite{wang2018driver} further proposed a model considering the complex mode choices among driving alone, public transit, using ride-sharing, and providing ride-sharing, which found that the capability of ride-sharing service to reduce vehicular traffic is not guaranteed and there won’t be a successful ride-sharing service without losing users of public transit when solo driving is faster but more expensive. \cite{ke2020pricing} modeled three key decision factors that determine the platform’s efficiency: trip fare, vehicle fleet size, and allowable detour time in monopoly, social optimum, and second-best solutions respectively. They proved that the fares in a ride-sharing market are lower than in a non-pooling market under certain conditions, and the demand-price elasticity of ride-pooling is higher than non-pooling.

Compared to equilibrium theory-based analysis, simulation-based studies focus more on the ride-sharing service itself and provide more flexibility in investigating the service performance and the complicated interactions between different factors. \cite{santi2014quantifying} proposed a shareability network to model the ride-sharing activity, where matching activity is simulated as an edge between nodes (trips) on the network. Using the New York City taxi data as an alternative demand information source, half of the trips were found to be shareable between two customers. \cite{alonso2017demand} further proposed a Ride-Trip-Vehicle (RTV) network to solve the intractability of sharing between more than two customers of \cite{santi2014quantifying} algorithm. Due to the model improvement, the percentage of shared trips was found to be at most 98\% with only 23\% taxis. \cite{yan2020quantifying} applied the shareability network model on the taxi trip in Shanghai, China, and quantified the ability of ride-sharing to reduce at least 15\% traffic emissions. Besides the network-based simulation model, many studies adopted the agent-based simulation method, where customers, drivers, and the platform were treated as three types of agents with their strategies and actions. \cite{ma2013t} defined the problem of dynamic ride-sharing and proposed a comprehensive large-scale ride-sharing framework. Based on a 3-month taxi dataset of 33,000 taxis in Beijing. Their framework served 25\% more trips and saved 13\% of the total vehicular distance. To be more realistic, further studies considered factors such as shifting activity and heterogeneity in accepted detour time \citep{lokhandwala2018dynamic}, and road congestion \citep{chen2024quantifying}. In the era of autonomous vehicles, \cite{fagnant2014travel} and \cite{fagnant2018dynamic} paid their attention to the potential benefit of the ride-sharing service provided by the shared autonomous vehicle (SAV) in their simulation and found one SAV can replace eleven conventional vehicles, which helps to relieve traffic congestion and can reduce total travel time even with a detour.

Although prior studies provided meaningful insights about the performance of ride-sharing, there are still a few research gaps. For equilibrium-based studies, the spatiotemporal disparities of ride-sharing can hardly be captured, while agent-based simulations often ignore the customers' price and detour elasticity. Therefore, this study integrates a survey that captures customers' price and detour elasticity and agent-based simulations to work out the economic and social performance of ride-sharing, revealing the potential barriers to ride-sharing development.

\section{Methodology}\label{sec: method}
\subsection{Dataset}

\subsubsection{Demand and road networks}

The demand datasets used in this study come from a giant TNC in China. The datasets cover travel orders in nine cities, including Shanghai, Shenzhen, etc., each containing 500,000 orders recorded from November 1st, 2015, to October 31st, 2016. Due to the privacy issue, these orders are uniformly downsampled from the original database but maintain the same spatial and temporal distributions. Each order record includes the coordinates (longitude and latitude) and timestamps of the origin and destination. These demand datasets can generally represent the demand patterns in the corresponding cities, which can be used to simulate the evolution of ride-sharing services. The vehicles' locations are initialized according to the spatial distributions of the demand, assigning more vehicles to areas with more orders.

\begin{figure}[!ht]
    \centering
    \includegraphics[width=0.99\linewidth]{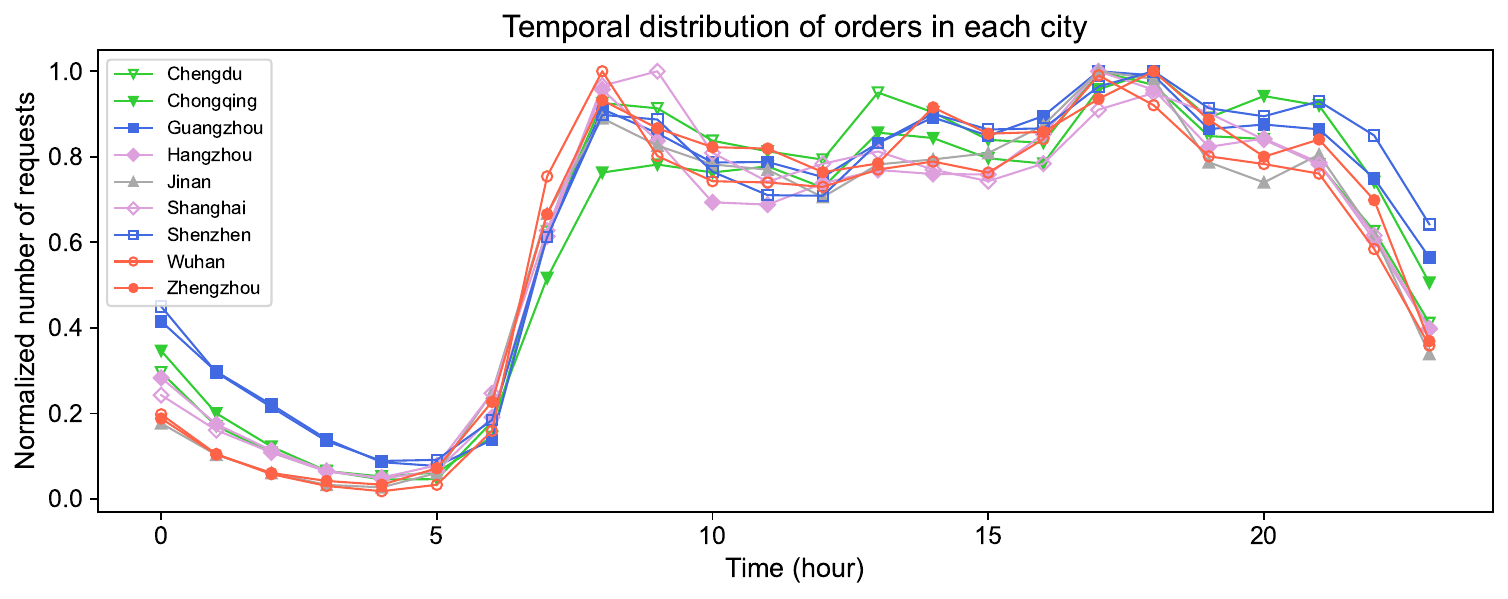}
    \caption{Temporal distribution of customers' requests in each city.}
    \label{fig: temporal_distribution}
\end{figure}

Figure \ref{fig: temporal_distribution} shows the temporal distribution of these datasets. The normalized number of requests is calculated as the ratio of the number of requests within the current period to the maximum number of requests across all periods. All orders in nine cities exhibit a similar temporal pattern. First, the morning and evening peak hours are around 8--9 am and 5--6 pm, respectively. In addition, the demand increases significantly after 7 am and decreases to a low level after 10 pm. During 8 am and 8 pm, the demand in all cities is relatively high.

The road networks of all cities are extracted from the Open Street Map \citep{OSM}. The road networks consist of nodes and links, representing intersections and roads, respectively. Only roads for vehicles are considered since this study focuses on ride-sharing services. Also, this study only considers the urban ride-sharing markets, including demand, supply, and rode networks, to reduce computational costs. This is reasonable because most of the orders in each city are concentrated in its urban area. Once the road networks are obtained, the origins and destinations of orders are relocated to their nearest nodes to facilitate calculating the routes on the road networks for vehicles. Given a pair of nodes, e.g., the origin and destination of an order, this study adopts the Dijksta algorithm \citep{dijksta1959note} to calculate the shortest route between them. It should be noted that traffic congestion is not considered in this study since it primarily focuses on the average performance of ride-sharing. Hence, vehicles are assumed to travel according to the shortest routes with a constant velocity.

\begin{figure}[!ht]
    \centering
    \includegraphics[width=0.9\linewidth]{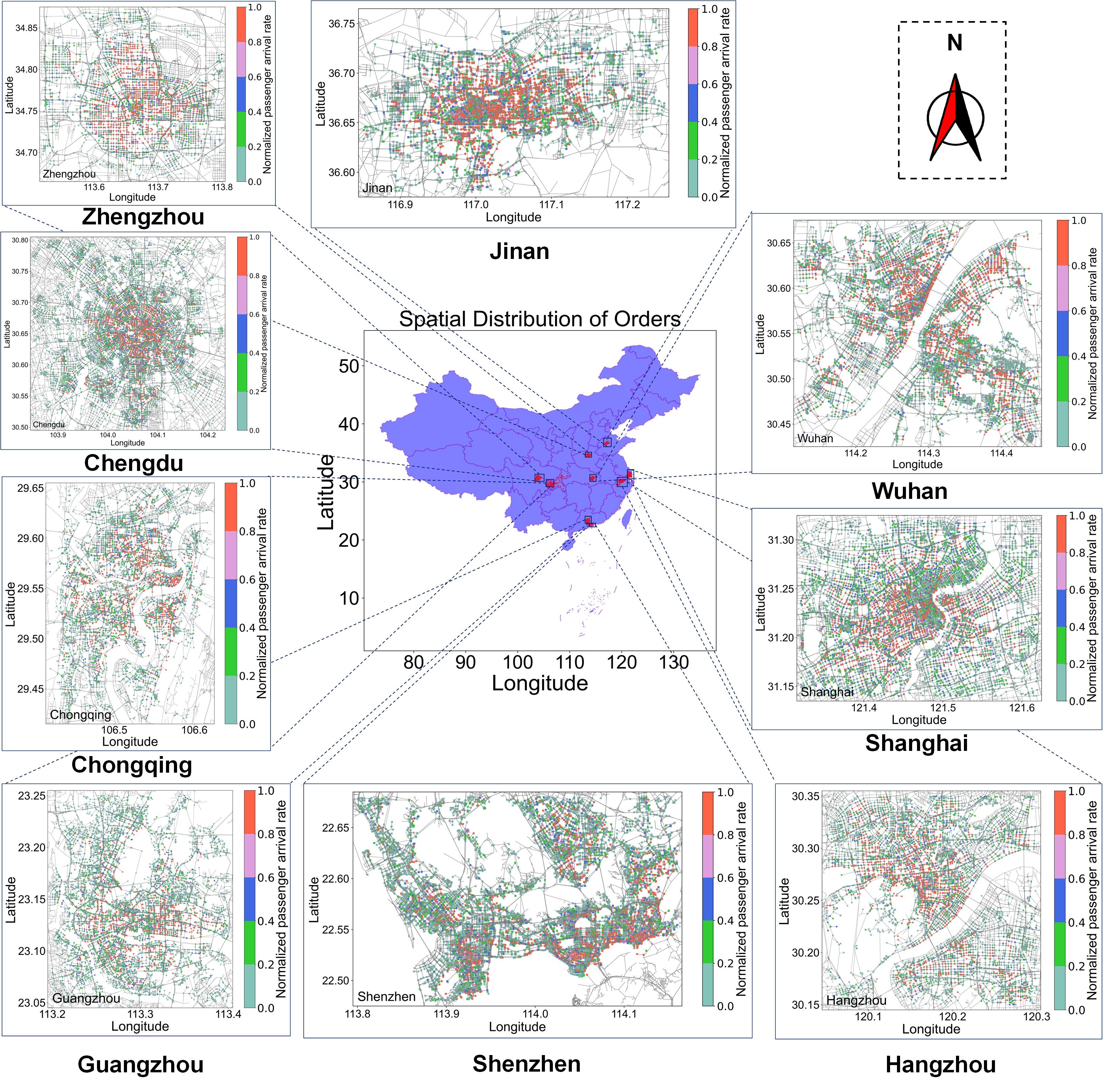}
    \caption{Spatial distribution of orders in each city.}
    \label{fig: spatial_distribution}
\end{figure}

Figure \ref{fig: spatial_distribution} illustrates the topology of the road networks and spatial distributions of demand in each city. The normalized customer arrival rate is calculated as the ratio of the order arrival rate at the current node to the maximum order arrival rate across all nodes. The structure of the road network in each city varies a lot. For example, the road networks in Chengdu, Zhengzhou, and Jinan are relatively uniform, with most orders concentrated in the central area. The road networks in the other cities are separated by a river, a few mountains, or even a sea, making their topology more complicated. These various road networks and spatial distributions of orders ensure the robustness of the experimental results as well as the corresponding conclusions of this study.

\subsubsection{Survey results}

This study conducted a questionnaire survey to calibrate customers' price and detour elasticity. It asked respondents to select a minimum acceptable discount for ride-sharing given a detour ratio. The detour ratio refers to the ratio of detour distance induced by ride-sharing to the original trip distance without ride-sharing. The questionnaire considers five detour ratios: 10, 20, 30, 40, and 50\%. Also, the discount considered in this study is the percentage of reduced price based on the original price. Given a detour ratio of ride-sharing, this study considers seven discount ratios: 5, 10, 15, 20, 30, 40, and 50\%. If a respondent declines all discount ratios, they will not participate in ride-sharing under this scenario. This consideration of the detour and discount ratio can represent various trips with different trip distances, avoiding distinguishing between long- or short-haul orders. This is reasonable because this study focuses on the average performance of ride-sharing, aiming to discover a few general findings on the revenue and social welfare of ride-sharing services.

\begin{figure}[!ht]
    \centering
    \subfigure[Response distribution]{\includegraphics[width=0.99\linewidth]{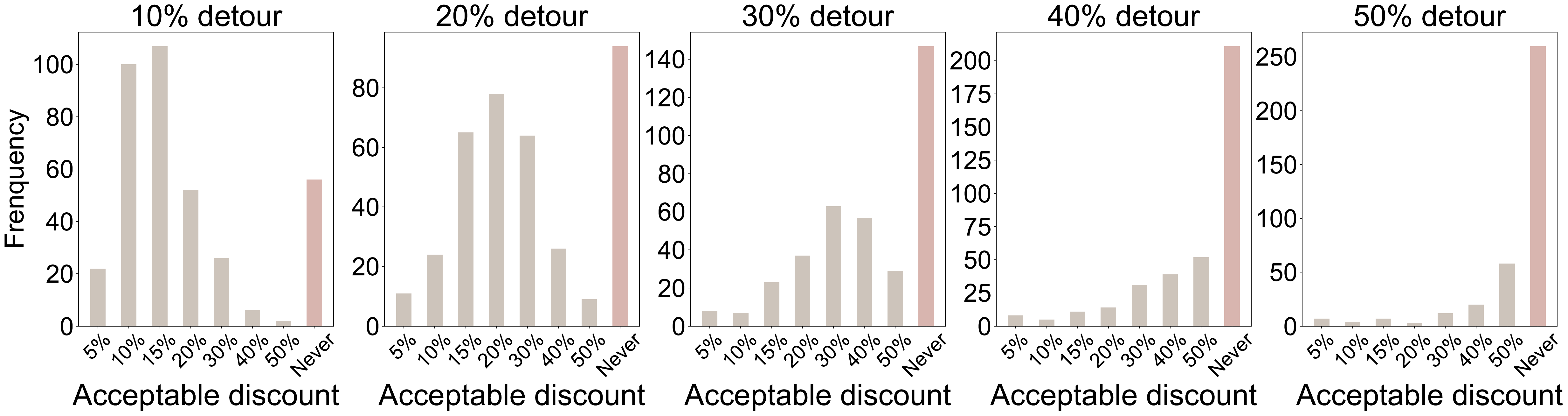}}
    \subfigure[Price and detour elasticity]{\includegraphics[width=0.99\linewidth]{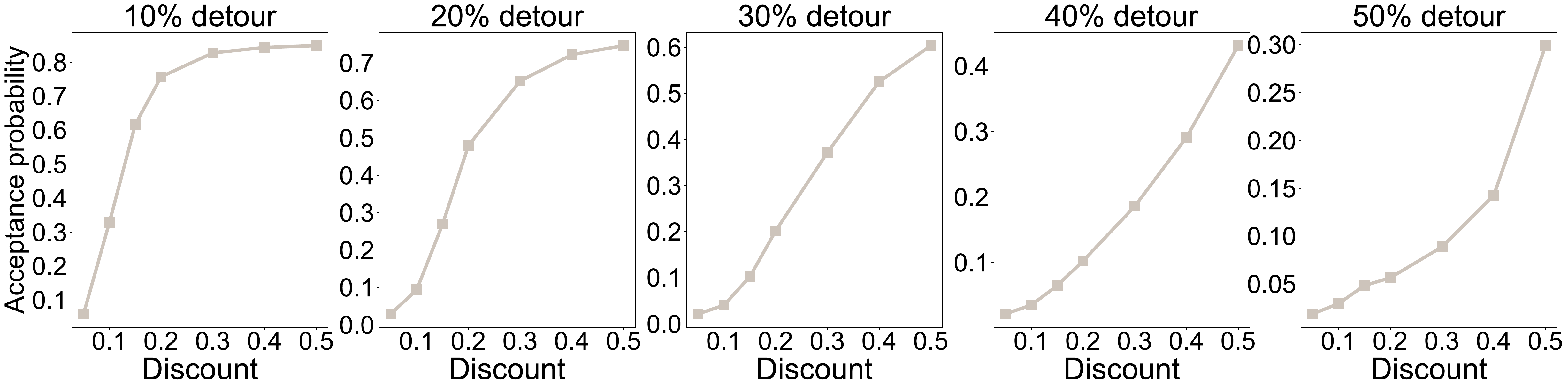}}
    \caption{Survey results and calibrated pricing and detour elasticity of customers.}
    \label{fig: survey_results}
\end{figure}

This study surveys several cities in China, including Shenzhen, Hong Kong, Chengdu, etc. Hence, the collected response can be representative of quantifying customers' price and detour elasticity. This study collected 402 effective questionnaires, each containing five scenarios corresponding to the five detour ratios considered. The survey results are shown in Figure \ref{fig: survey_results} (a). When the detour ratio is low (e.g., 10\%), many respondents are willing to participate in ride-sharing with a relatively low discount ratio (e.g., 10--20\%). However, most respondents decline ride-sharing when the detour ratio exceeds 30\%. The explicit price elasticity of customers involving different detour ratios is approximated using the frequency of respondents choosing different discount ratios. The calibrated price and detour elasticity of customers are shown in Figure \ref{fig: survey_results} (b).

A few interesting insights can be inferred from the calibration results. First, when the induced detour ratio is low, customers are more price-sensitive. For example, given a detour ratio of 10\%, if the offered discount ratio is 5\%, then no more than 10\% customers are willing to participate in ride-sharing, while around 60\% customers will accept it if the offered discount ratio is 15\%. In this way, a relatively low discount ratio can attract many customers to ride-sharing. This provides a valuable reference for designing the pricing strategy for hot areas with high demand since the detour ratio can be low if the demand is dense. Another interesting finding is if the detour ratio is high, a TNC needs to offer significantly low prices to maintain ride-sharing services. For instance, if the detour ratio is 50\%, then no more than 30\% customers will participate in ride-sharing services even offered with a 50\% discount. This situation is similar to the ride-sharing services in suburban areas that require long detour distances due to the relatively sparse demand. In this situation, subsidies from TNCs or governments are necessary to maintain the ride-sharing services, increasing traffic accessibility in suburban areas and enhancing transportation equity across different regions. We further discuss these insights in Section \ref{sec: discussion}.

\subsection{Ride-sharing mechanism}

This study considers an online ride-hailing platform operated by a TNC in each city. The platform connects customers and vehicles. This study considers two scenarios: (1) the platform provides only solo-hailing services, and (2) the platform provides mixed solo-hailing and ride-sharing services. The revenue and social welfare (e.g., carbon emissions) of these two scenarios are compared to quantify the pros and cons of introducing ride-sharing services. Under the first scenario, all customers are assumed to participate in the solo-hailing service since this study considers only one platform. While customers may choose solo-hailing or ride-sharing according to the upfront prices and maximum detour ratio.

\begin{figure}[!ht]
    \centering
    \includegraphics[width=0.7\linewidth]{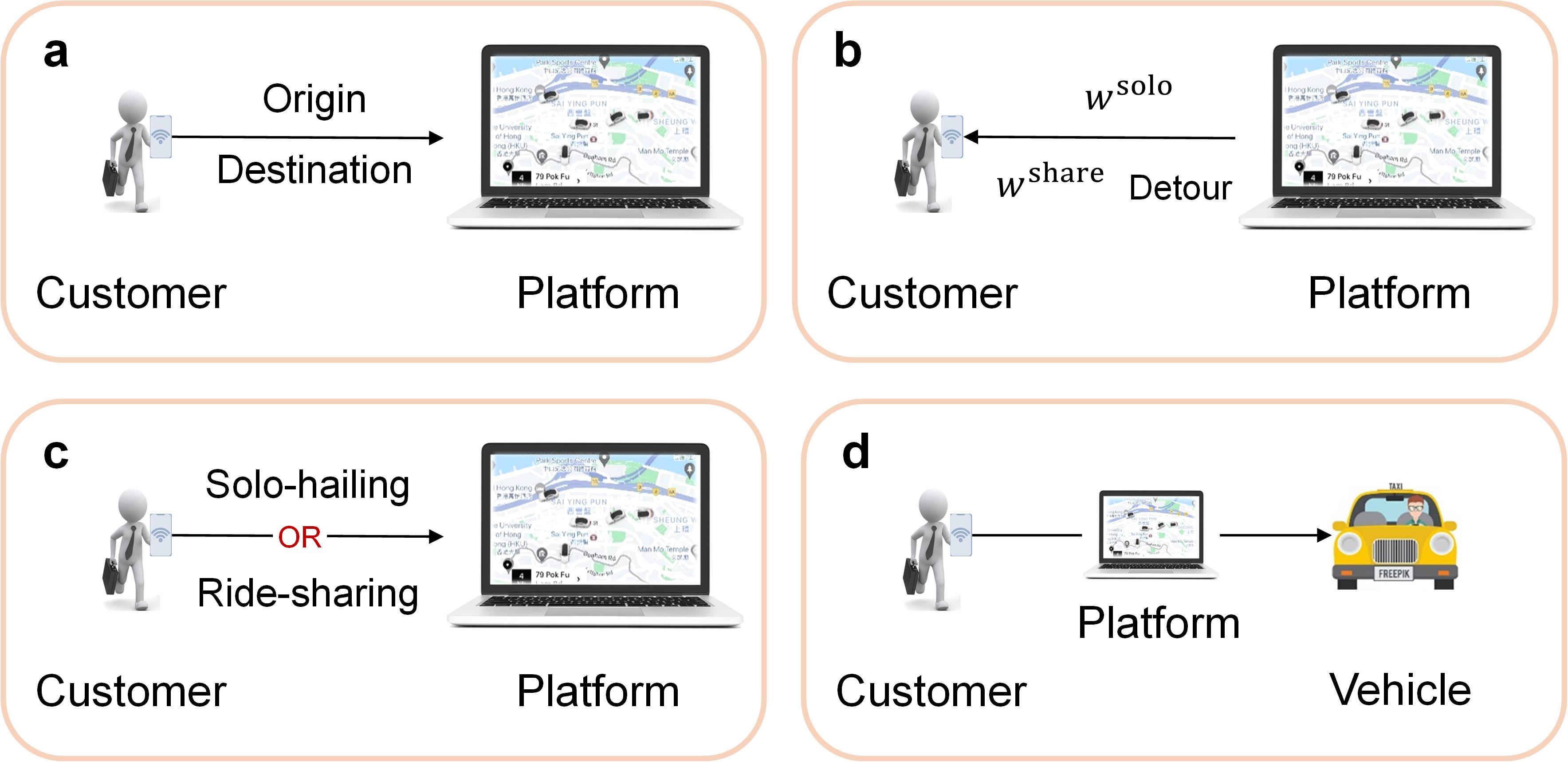}
    \caption{Illustration of upfront pricing mechanism of mixed solo-hailing and ride-sharing services.}
    \label{fig: upfront_pricing}
\end{figure}

The upfront pricing mechanism of the mixed services is shown in Figure \ref{fig: upfront_pricing}. First, the customers submit their origins and destinations to the platform. Then, the platform offers upfront prices, including solo-hailing and ride-sharing prices, to the customers. In real-world applications, a platform may adjust pricing strategies according to the real-time demand-to-supply ratios to maximize revenue. However, this study does not aim to optimize the platform's revenue and just assumes the platform adopts a consistent strategy across the entire horizon. Specifically, the solo-hailing price is calculated as the basic trip fare plus a fare proportional to the trip distance, and the ride-sharing price is a discount on the solo-hailing price. The upfront solo-hailing and ride-sharing prices (denoted as $w^{\text{solo}}$ and $w^{\text{share}}$) are calculated as follows:

\begin{equation}
    w^{\text{solo}} = k_1 + k_2 d,
\end{equation}

\begin{equation}
    w^{\text{share}} = (1-\theta)w^{\text{solo}},
\end{equation}
where $k_1$ and $k_2$ respectively denote the basic fare and additional fare per unit distance, $d$ the trip distance, and $\theta$ the offered discount for ride-sharing. In addition, the platform offers a maximum detour guarantee of the ride-sharing service to the customers, that is, the platform promises the detour ratio will not exceed the guaranteed value (e.g., 0.3) if a customer chooses ride-sharing. Subsequently, based on the offered upfront prices and detour guarantee, the customers choose between solo-hailing and ride-sharing. Finally, the platform assigns vehicles to serve the customers. In particular, if a customer chooses the solo-hailing service, then the platform can only match the customer with an idle vehicle. However, if a customer decides to participate in ride-sharing, then the platform can assign the customer to an idle or partially occupied vehicle.

\begin{figure}[!ht]
    \centering
    \includegraphics[width=0.7\linewidth]{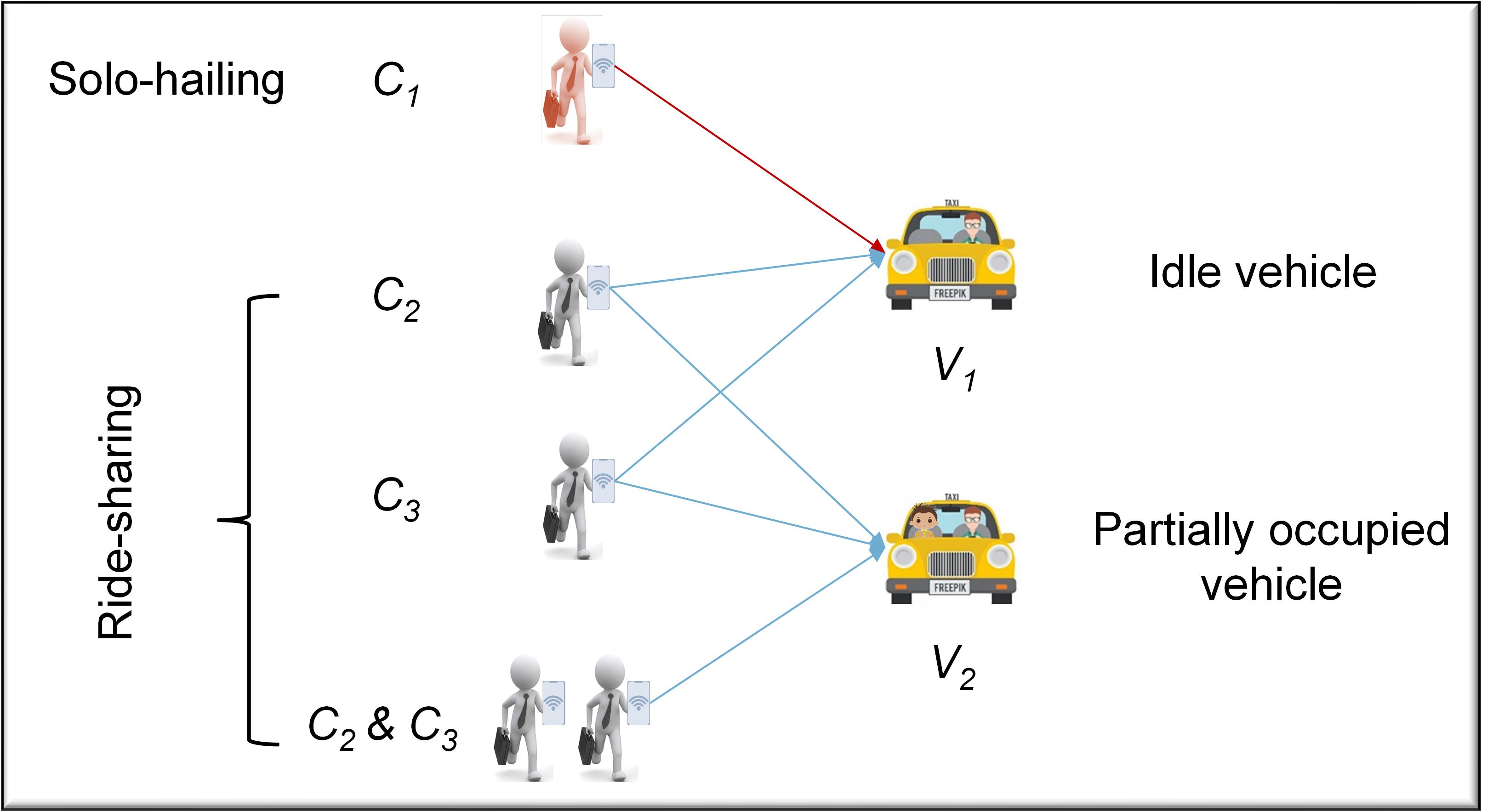}
    \caption{Illustration of customer-vehicle assignment.}
    \label{fig: matching}
\end{figure}

The platform assigns customers to vehicles using the batch-matching mechanism, i.e., the platform matches a batch of vehicles and customers every once in a while. The customer-vehicle assignment algorithm used in this study is similar to that proposed by \cite{alonso2017demand}, but can handle solo-hailing and ride-sharing assignments simultaneously. Figure \ref{fig: matching} illustrates the customer-vehicle assignment. In particular, the customer $C_1$ chooses the solo-hailing service and can only be assigned to the idle vehicle $V_1$. The customers $C_2$ and $C_3$ are willing to participate in ride-sharing, they can be matched with the idle vehicle $V_1$ as well as the partially occupied vehicle $V_2$. Moreover, before the assignment, the platform determines which customers who choose the ride-sharing service can be pooled together according to their itineraries. Specifically, the platform first calculates the shortest route for a vehicle to sequentially visit the origins and destinations of these customers. Then, the platform checks whether the shortest route satisfies the maximum detour ratio constraint for all customers. If yes, these customers can share trips; otherwise, they cannot be pooled together. Please refer to \cite{chen2024quantifying} for the details of the matching process. If multiple customers can be pooled, they are merged into an additional trip, e.g., $C_2\&C_3$ in Figure \ref{fig: matching}. At the same time, each customer in the merged trip can also be an independent trip, e.g., $C_2$ and $C_3$ in Figure \ref{fig: matching}.

To be more specific in implementation, the mathematical formulation of the above assignment is introduced as follows. Let $R$ and $J$ denote the batch of customers and vehicles, respectively. Let $I$ be the trips after pooling customers. Remind that each trip may contain one or multiple customers. In addition, let $x_{ij} \in \{0, 1\}, \forall i \in I, j \in J$ be the decision variable that determines whether trip $i$ and vehicle $j$ are matched. Let $u_{ij}$ denote the utility of assigning trip $i$ to vehicle $j$, which can be calculated as follows:

\begin{equation}
    u_{ij} = w_i - c_{ij},
\end{equation}
where $w_i$ denotes the price for the trip $i$ and $c_{ij}$ the cost for vehicle $j$ to serve trip $i$. Therefore, the customer-vehicle assignment can be formulated as an integer linear problem (ILP), as follows:

\begin{equation}
    \max_{x_{ij}} u_{ij}x_{ij}
\end{equation}
s.t.,

\begin{equation}\label{con: 1}
    \sum_{i \in I} x_{ij} \leq 1, \quad \forall j \in J,
\end{equation}

\begin{equation}\label{con: 2}
    \sum_{j \in J}\sum_{r\in i, i\in I} x_{ij} \leq 1, \quad \forall r \in R,
\end{equation}

\begin{equation}
    x_{ij} \in \{0, 1\}, \quad \forall i \in I, j \in J.
\end{equation}
Constraints \eqref{con: 1} ensure each vehicle can be assigned at most one trip, and constraints \eqref{con: 2} ensure each customer can be matched with at most one vehicle. After the assignment, the vehicles matched with customers travel to pick up customers along the shortest routes. In addition, this study relocates those idle vehicles without assignment to the positions of waiting customers according to the repositioning algorithm provided by \cite{alonso2017demand}.

\subsection{Agent-based simulation}

This study achieves the above-mentioned mechanism on an agent-based simulation platform. The architecture of the simulation platform is introduced in \cite{chen2024quantifying}. This study further enriches the functionality of the platform by implementing the upfront pricing mechanism and joint assignment of solo-hailing and ride-sharing services. The discount and detour guarantee for ride-sharing are given before the simulation and remain the same across the entire simulation horizon. The customers are assumed to be homogeneous and have the same probability to accept or decline ride-sharing for the given discount and maximum detour ratio. In addition, each order is assumed to contain only one customer. Once the customers make their decisions, they are assumed not to switch to another service. If a customer has not been matched with any vehicles for a while (e.g., 10 minutes), they will abandon their requests and leave the platform. The vehicles are assumed to fully comply with the platform's dispatch. At the end of simulations, all metrics, e.g., revenue and carbon emissions, are calculated and recorded for further analysis.

\subsection{Measurement}

\subsubsection{System performance}

This study considers two measurements to quantify the system performance: revenue and social welfare. The former is the most concerned metric of TNCs and drivers, indicating the economic sustainability of ride-sharing services. If introducing ride-sharing cannot increase the revenue compared with a pure solo-hailing service, then drivers or TNCs do not have a strong motivation to provide ride-sharing services or even improve the service quality. Regarding social welfare, this study extends it to a broader perspective, including the system efficiency, carbon emissions, and customers' waiting time. These metrics can comprehensively measure the impact brought by ride-sharing services.

In this study, the revenue distribution between TNCs and drivers is not considered. Hence, The revenue refers to the total trip fare charged from those customers who have been served. Social welfare includes five metrics, as follows:

\begin{itemize}
    \item \textbf{Service rate} refers to the ratio of the number of served customers to the total number of customers. This metric measures the system's efficiency given a fleet of vehicles.
    
    \item \textbf{Average number of scheduled requests} refers to the average number of customers each vehicle has been scheduled, including customers already onboard and those scheduled to be picked up in the future. This metric is calculated as the mean value of the numbers recorded after every assignment across the entire simulation. This metric measures the average vehicle occupancy rate of the vehicles in the system.
    
    \item \textbf{Delivery emission factor} refers to the average CO$_2$ emissions required for delivering one kilometer of customers' requests. The CO$_2$ emissions are calculated using the COPERT model developed by Europe Environment Agency \citep{COPERT2020}. The details of CO$_2$ calculate can be found in \cite{chen2024quantifying}. Remind that the carbon emissions for relocating idle vehicles are also considered when calculating the emission factor. This metric considers the total carbon emissions and delivered customers, which can be representative of measuring the carbon emissions of the solo-hailing or mixed systems.
     
    \item \textbf{Matching time} refers to the time between a customer choosing a service and the platform matching the customer with a vehicle.

    \item \textbf{Pickup time} refers to the time between the platform matching a customer with a vehicle and the vehicle picking up the customer.
\end{itemize}

\subsubsection{Ride-sharing efficiency}

\begin{figure}[!ht]
    \centering
    \includegraphics[width=0.7\linewidth]{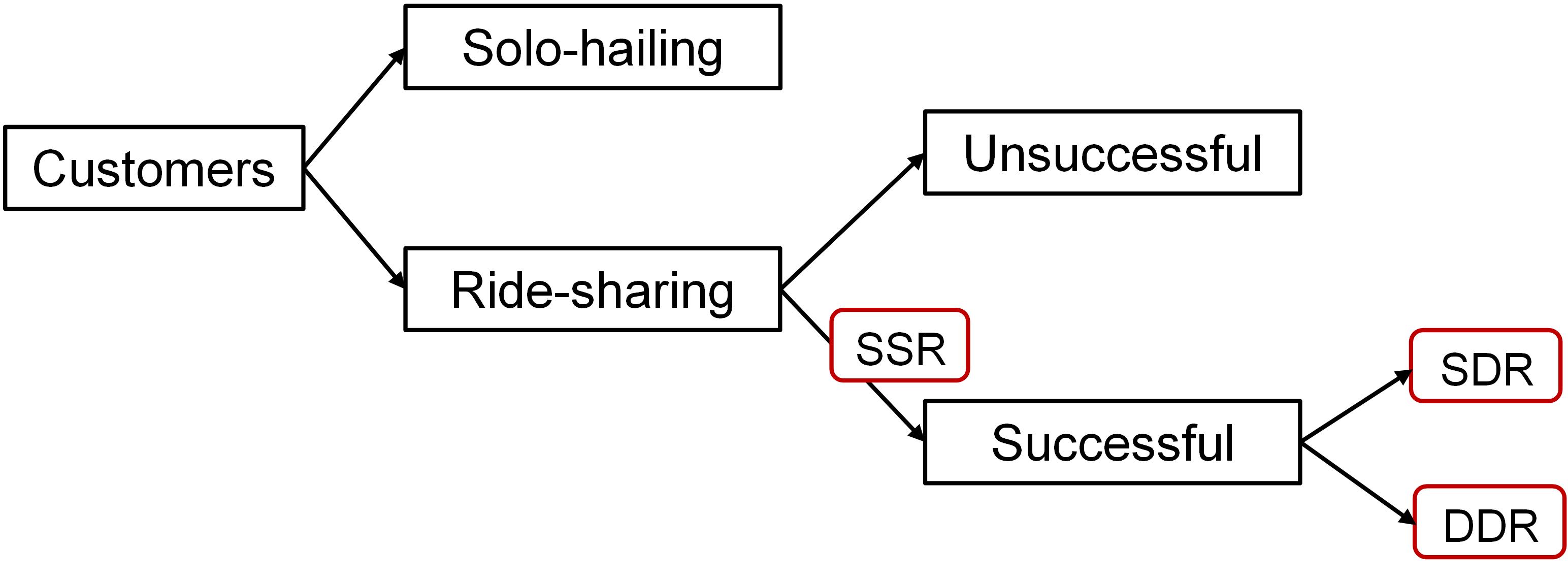}
    \caption{Illustration of the metrics to measure the performance of ride-sharing.}
    \label{fig: ST_metrcis}
\end{figure}

This study considers three metrics to measure the efficiency of the ride-sharing service in the mixed system, as follows:

\begin{itemize}
    \item \textbf{Successful sharing ratio (SSR)} refers to the ratio of the number of customers sharing their trips with others successfully to the number of customers choosing the ride-sharing service.

    \item \textbf{Shared distance ratio (SDR)} refers to the ratio of customers' shared distance with others to their original trip distance without ride-sharing.

    \item \textbf{Detour distance ratio (DDR)} refers to the ratio of customers' detour distance induced by ride-sharing to their original trip distance.
\end{itemize}
The relationship of the three metrics is shown in Figure \ref{fig: ST_metrcis}. The platform should increase the SSR and SDR and decrease the DDR as much as possible to enhance the efficiency of ride-sharing. Hence, these three metrics are representative of measuring the efficiency of ride-sharing.

\begin{figure}[!ht]
    \centering
    \includegraphics[width=0.5\linewidth]{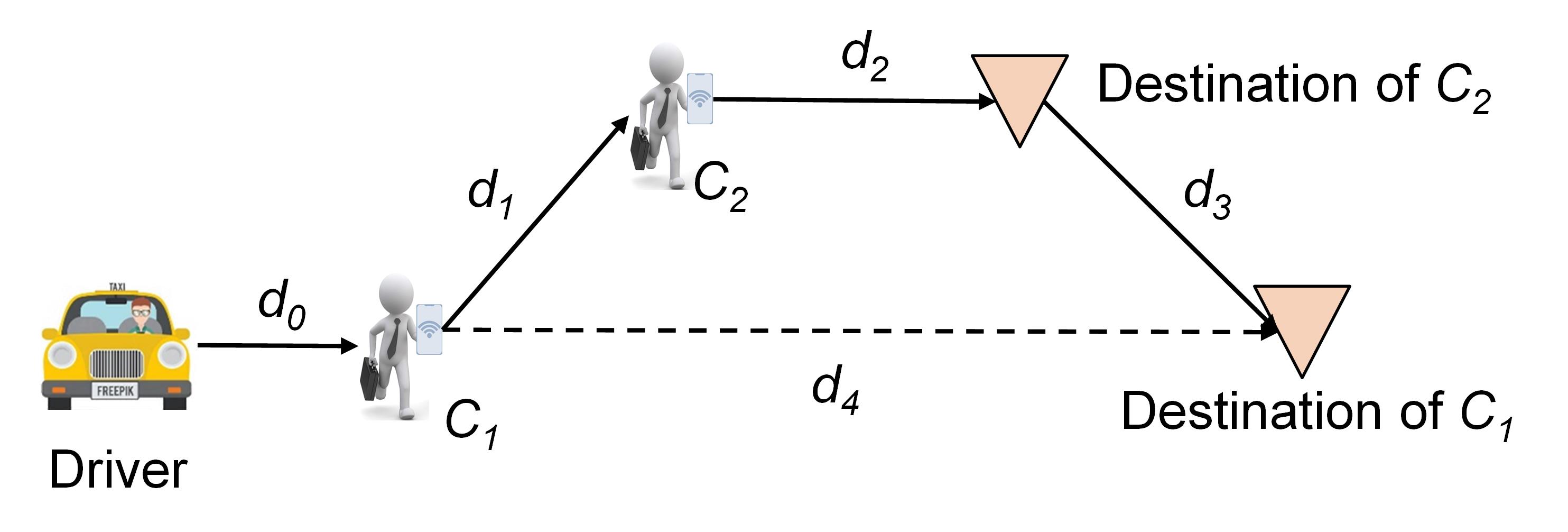}
    \caption{Illustration of shared, detour, and saved distance calculation. Solid arrows represent the route of the shared trip, and $d_4$ and $d_2$ denote the original trip distance of $C_1$ and $C_2$, respectively.}
    \label{fig: shared_dis_calculation}
\end{figure}

Figure \ref{fig: shared_dis_calculation} illustrates the calculation of shared, detour, and saved distances. Specifically, the shared distance for both customers is $d_2$. The detour distances for $C_1$ and $C_2$ are $d_1 + d_2 + d_3 - d_4$ and 0, respectively. Regarding the saved distance, it is calculated as the difference between two customers' original trip distances and the actual distance of the shared trip. Hence, the saved distance for pooling $C_1$ and $C_2$ is $(d_2 + d_4) - (d_1 + d_2 + d_3) = d_4 - d_1 - d_3$. Moreover, the saved carbon emissions can be calculated based on the saved distance by using the COPERT model \citep{COPERT2020}.

\section{Experiments and results}\label{sec: exp}

\subsection{Experiment setup}

This study downsamples orders from the datasets to facilitate simulations since this study conducts experiments in nine cities. Specifically, this study adopts around 20\% of the original orders in each city. In addition, only customers and road networks in the urban area are considered during the simulation. The simulation horizon is from 6 am to 12 pm. The request arrival rate increases significantly during this period, which can capture the demand patterns in each city. In addition, we find that the experimental results exhibit trends similar to those of the 18-hour simulations reported by \cite{chen2024deconomic} in Section \ref{sec: overall_results}, verifying that the simulation horizon adopted is reasonable and effective. After the end of the simulation horizon, this study simulates another 45 minutes, ensuring that all scheduled customers are delivered to their destinations. The request arrival rates and urban areas of nine cities are listed in Table \ref{tab: arrival_rate}.

\begingroup
\setlength{\tabcolsep}{6pt} 
\renewcommand{\arraystretch}{1} 
\begin{table}[!ht]
\caption{Customers' arrival rate, average distance, and the urban area in each city.}
\begin{center}
\begin{tabular}{ l c c c}
\toprule
City & Arrival rate & Average distance & Urban area \\
 & (\#/h) & (km) & (km$^2$) \\
\midrule
Chengdu   & 1624 & 8.0 & 1119 \\ 
Chongqing & 1604 & 7.0 & 495 \\ 
Guangzhou & 2260 & 6.3 & 455 \\ 
Hangzhou  & 2186 & 6.0 & 535 \\ 
Jinan     & 2274 & 6.1 & 715 \\ 
Shanghai  & 2100 & 6.3 & 572 \\ 
Shenzhen  & 1598 & 7.7 & 823 \\ 
Wuhan     & 1970 & 6.6 & 844 \\ 
Zhengzhou & 1964 & 6.2 & 570 \\

\bottomrule
\end{tabular}
\label{tab: arrival_rate}
\end{center}
\end{table}

The basic trip fare is 4 CNY, and the fare proportional to trip distance is 3 CNY/km. The minimum trip fare for a customer is 10 CNY, even if their trip distance is very short, which aligns with the pricing configuration of TNCs in China. This study considers discount ratios of 0, 0.1, 0.15, 0.2, 0.3, and 0.4. In particular, 0 discount represents the pure solo-hailing service. In addition, the guaranteed detour ratios considered in this study are 20, 30, and 40\%, respectively. The fleet size is 500 in each city, with a maximum capacity of 2 customers. Hence, the scenarios with different detour ratios are denoted as C2-D20, C2-D30, and C2-D40, respectively.

\subsection{Overall results}\label{sec: overall_results}

The experimental results in three different scenarios, C2-D20, C2-D30, and C2-D40, are similar. Therefore, for simplicity, only the details of C2-D30 are analyzed and discussed in this section, with the results of the other scenarios presented in \ref{seca: exp_results}.

Figure \ref{fig: revenue_sr} shows the average revenue and service rate of each city's pure solo-hailing and mixed systems. Compared with the pure solo-hailing service (0 discount), except in Jinan, the revenue of the mixed services decreases as the discount increases in all cities. In particular, when the discount exceeds 0.2, the revenue can be significantly decreased. This indicates that introducing ride-sharing into solo hailing may reduce the revenue of TNCs or drivers without carefully designing pricing strategies. In this way, drivers may not be willing to serve ride-sharing orders, restricting the growth of the market share of ride-sharing services. However, the average service rate increases as the discount ratio increases in all cities. This indicates introducing ride-sharing can improve the system's efficiency, accommodating more requests given a fleet of vehicles. In particular, when the discount increases from 0.1 to 0.2, the increase in the service rate is the most significant. While the discount exceeds 0.2, the marginal benefit of increasing the ride-sharing discount is insignificant. The vehicle occupancy rate is relatively high when the discount ratio is 0.2. Hence, even if more customers are willing to participate in ride-sharing, the platform finds it more difficult to pool these customers. As a result, the service rate of customers cannot dramatically increase when the discount for ride-sharing further increases. In this context, TNCs do not need to offer a high discount ratio for customers as it cannot dramatically increase the system's efficiency, but it may reduce revenue significantly.

\begin{figure}[!ht]
    \centering
    \subfigure[Revenue]{\includegraphics[width=0.4\linewidth]{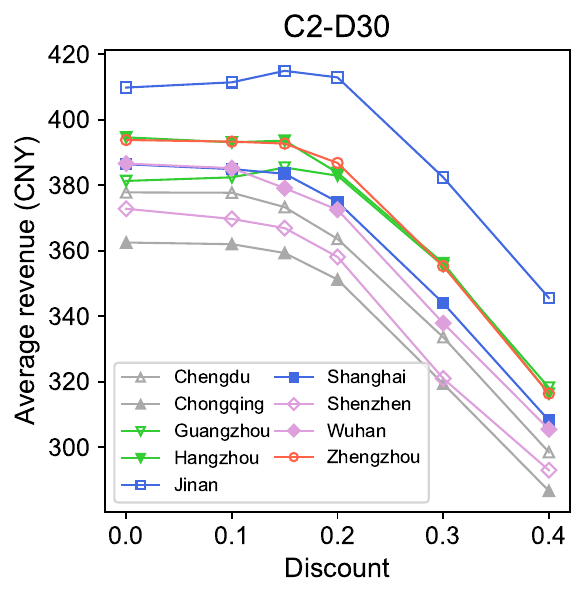}}
    \subfigure[Service rate]{\includegraphics[width=0.4\linewidth]{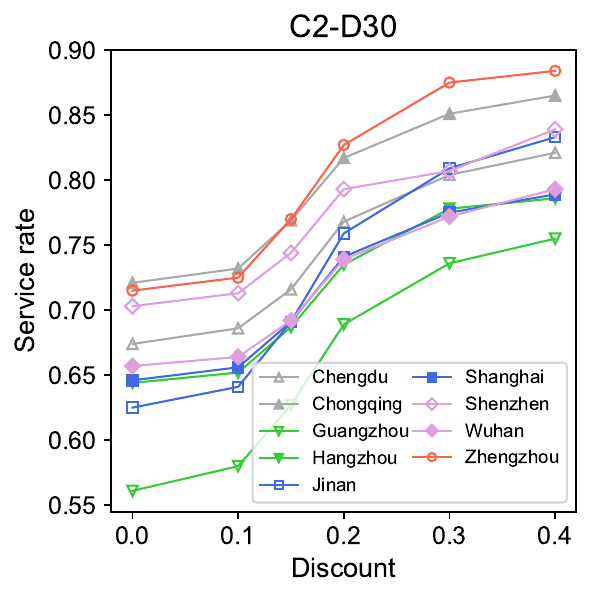}}
    \caption{Experiment results on the average revenue and service rate.}
    \label{fig: revenue_sr}
\end{figure}

Figure \ref{fig: emission} shows the experimental results on carbon emissions and vehicle occupancy rate. Introducing ride-sharing can decrease the required CO$_2$ emissions for delivering one kilometer of customers' requests. This is because ride-sharing enables two customers to share their trips, reducing the required carbon emissions to satisfy the same customers' requests. Also, ride-sharing can increase the vehicle occupancy rate as one vehicle can carry two customers on a trip. Remind that idle vehicles are also considered when calculating the vehicle occupancy rate. In the pure solo-hailing system (with 0 discount), the average number of scheduled requests of each vehicle is around 0.8, which means approximately 20\% vehicles are idle on average across the entire horizon. While ride-sharing can enhance the vehicle utilization rate by enabling a vehicle to accommodate two customers. Similarly, when the discount is high, e.g., over 0.2, the marginal benefits of increasing the discount for ride-sharing are insignificant in terms of carbon emission reductions and vehicle occupancy rate improvement. Therefore, it is not necessary for TNCs to offer a very high discount to attract more customers to participate in ride-sharing.

\begin{figure}[!ht]
    \centering
    \subfigure[Carbon emissions]{\includegraphics[width=0.4\linewidth]{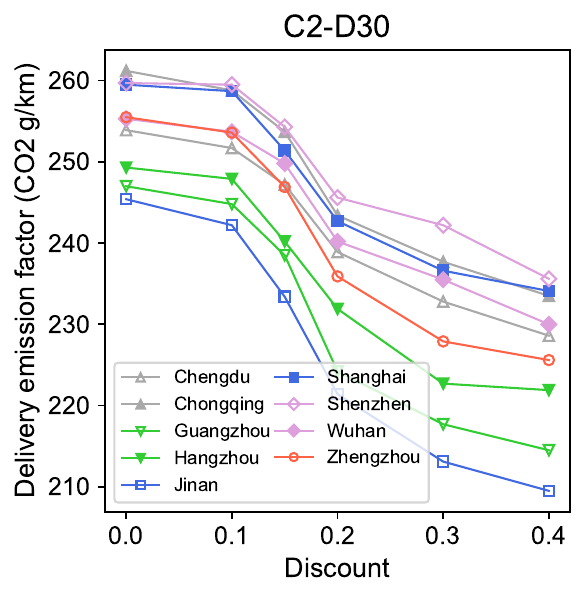}}
    \subfigure[Vehicle occupancy rate]{\includegraphics[width=0.4\linewidth]{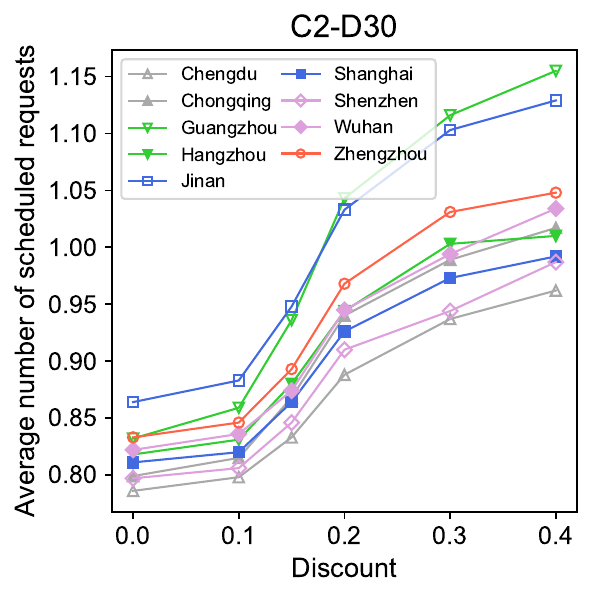}}
    \caption{Experiment results on the carbon emissions and vehicle occupancy rate.}
    \label{fig: emission}
\end{figure}

Customers' average matching and pickup times are shown in Figure \ref{fig: pickup_time}. Compared with the pure solo-hailing service, introducing ride-sharing can reduce customer matching time, indicating the platform can match customers with vehicles faster in the system with mixed services. Customers need to wait for idle vehicles in the pure solo-hailing system, while they can share vehicles with others in the mixed system. Therefore, the available vehicle arrival rate is higher in the mixed system, reducing customers' waiting times for assignment. In addition, introducing ride-sharing almost does not change the pickup time of customers, indicating the platform does not match customers with vehicles a longer distance away on average. These results demonstrate that ride-sharing can reduce customers' waiting times, which is significant for customers' choices. In real-world applications, customers may shift their choices to ride-sharing due to the shorter waiting time, inducing a higher demand for ride-sharing services. However, how to estimate the matching time for customers choosing solo-hailing or ride-sharing services can be a challenging task, and we leave it for future research. Nevertheless, these results are still valuable for quantifying the advantages and disadvantages of introducing ride-sharing.

\begin{figure}[!ht]
    \centering
    \subfigure[Matching time]{\includegraphics[width=0.4\linewidth]{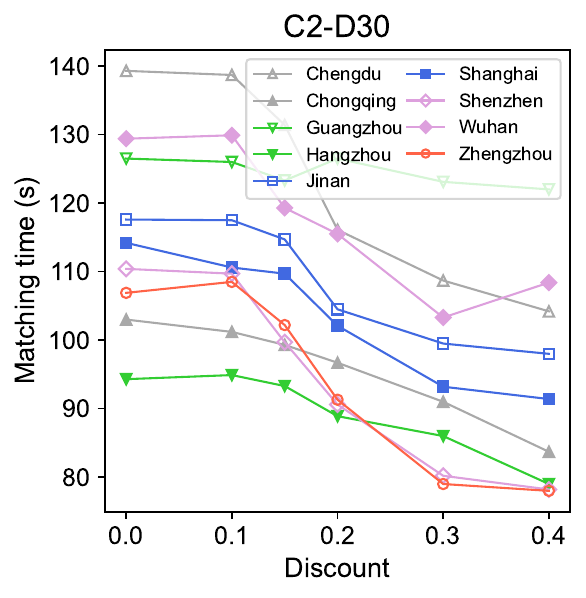}}
    \subfigure[Pickup time]{\includegraphics[width=0.4\linewidth]{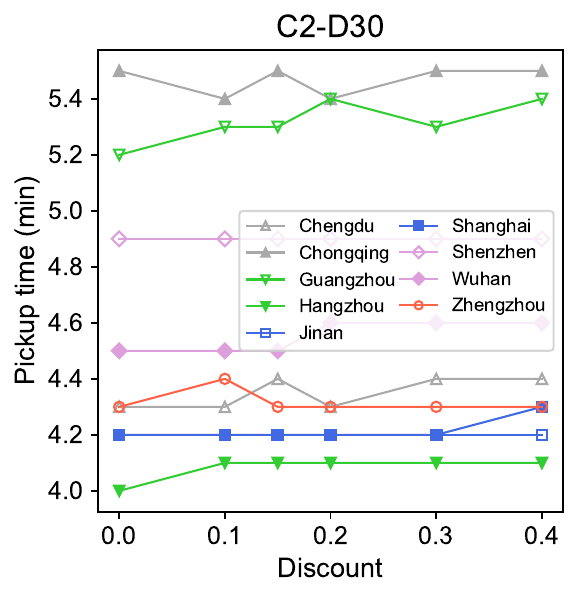}}
    \caption{Experiment results on customers' average matching and pickup times.}
    \label{fig: pickup_time}
\end{figure}

To further quantify the explicit implications of ride-sharing on the performance of the mixed system, we plot the increased or decreased percentage of five metrics with a 0.2 discount rate, as shown in Figure \ref{fig: percentage}. In particular, the waiting time consists of customers' matching and pickup times. On the one hand, introducing ride-sharing can increase customers' service rate by 15.7\%, decrease CO$_2$ emissions by 7.1\%, increase the vehicle occupancy rate by 16.7\%, and decrease customers' waiting times by 2.6\%. Therefore, from the perspective of the whole society, ride-sharing should be encouraged and its market share can be further enlarged. On the other hand, however, ride-sharing reduces the revenue by 2.3\% compared with the pure solo-hailing service. The revenue loss can be significant or even unacceptable for TNCs or drivers, affecting their proactivity to participate in ride-sharing services. This dilemma of ride-sharing is a key issue that TNCs and urban managers should consider when designing strategies and policies for ride-sharing services. One possible solution to avoid revenue loss is to design a dynamic pricing strategy for ride-sharing. But this is out of the scope of this study, and we leave it for future research. Another method is to take advantage of government subsidies to make ride-sharing a public service. We will discuss this point thoroughly in Section \ref{sec: ride-sharing-public}.

\begin{figure}[!ht]
    \centering
    \includegraphics[width=0.65\linewidth]{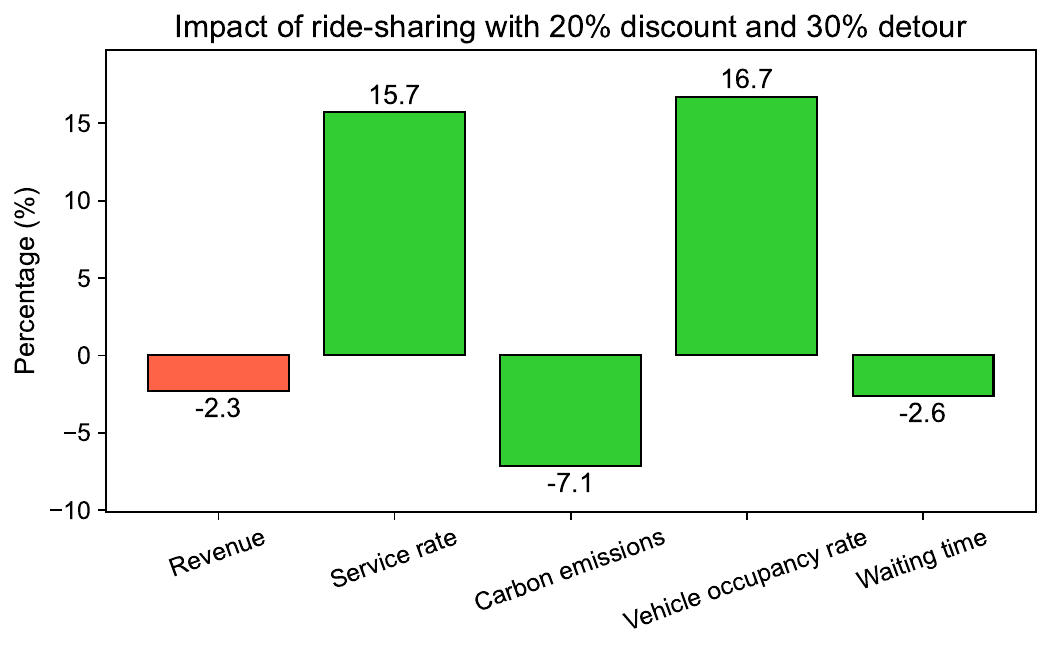}
    \caption{Impact of ride-sharing on revenue and social welfare with 20\% discount and 30\% detour.}
    \label{fig: percentage}
\end{figure}

\subsection{Spatiotemporal analysis}\label{sec: st_analysis}

To further figure out the implications of introducing ride-sharing into solo-hailing services, this study divides the study area in each city into three types according to their spatiotemporal characteristics: cold area, normal area, and hot area. Specifically, this study splits the time horizon into 12 slots, with one slot representing half an hour. Also, the study area in each city is divided into 100 zones equally. Consequently, the demand-to-supply ratio in each zone within each time slot can be calculated. This study adopts a dimensionless variable to measure the demand-to-supply, as follows:

\begin{equation}\label{eq: load}
    x = \frac{\lambda \Bar{d}}{Nv},
\end{equation}
where $x$ denotes the normalized load, $\lambda$ the customer arrival rate, $\Bar{d}$ the average distance of the customers' requests, $N$ the number of vehicles, and $v$ the travel speed of vehicles. In Equation \eqref{eq: load}, $\lambda \Bar{d}$ and $Nv$ respectively represent the demand and supply, and thus, the normalized load $x$ can measure the demand-to-supply ratio of each zone within a time slot. If $x$ equals 1, the demand and supply are balanced, and all customers can be accommodated \citep{molkenthin2020scaling}. Hence, this study defines a zone within a time slot as a cold area or hot area if its corresponding normalized load is smaller than 0.5 ($x < 0.5$) or larger than 2 ($x > 2$); otherwise, it is defined as a normal area when $0.5 \leq x \leq 2$.

Figure \ref{fig: ST_distribution} shows the performance of the ride-sharing service in three kinds of areas in each city. The SSR is relatively high in hot areas but significantly lower in cold areas. For example, even with a 0.4 discount for ride-sharing, the SSR of customers is around 40\% in cold areas, demonstrating that the platform cannot pool most customers choosing ride-sharing but still charges them a discounted price. Even in hot areas, the platform must offer a relatively high discount to maintain a relatively high SSR by attracting more customers to ride-sharing. These results indicate the dilemma of introducing ride-sharing: if the platform would like to improve the SSR, it needs to offer a higher discount, which in turn results in a loss of revenue. Regarding the SDR, customers have a slightly higher SDR in hot areas than in normal and cold areas, indicating the platform can pool customers and match them with vehicles faster in hot areas. For the customers choosing ride-sharing in cold areas, however, it may take some time for the platform to seek customers to share trips with them due to the low demand-to-supply ratio, leading to a relatively low SDR. As for the DDR, customers choosing ride-sharing in all areas have a similar DDR. This is because the platform only guarantees the maximum detour ratio and can match new customers with partially occupied vehicles en route. In this way, the platform will try to pool customers together as long as the induced detour ratio does not exceed the guaranteed, leading to a similar DDR of customers choosing ride-sharing across all areas.

\begin{figure}[!ht]
    \centering
    \includegraphics[width=0.99\linewidth]{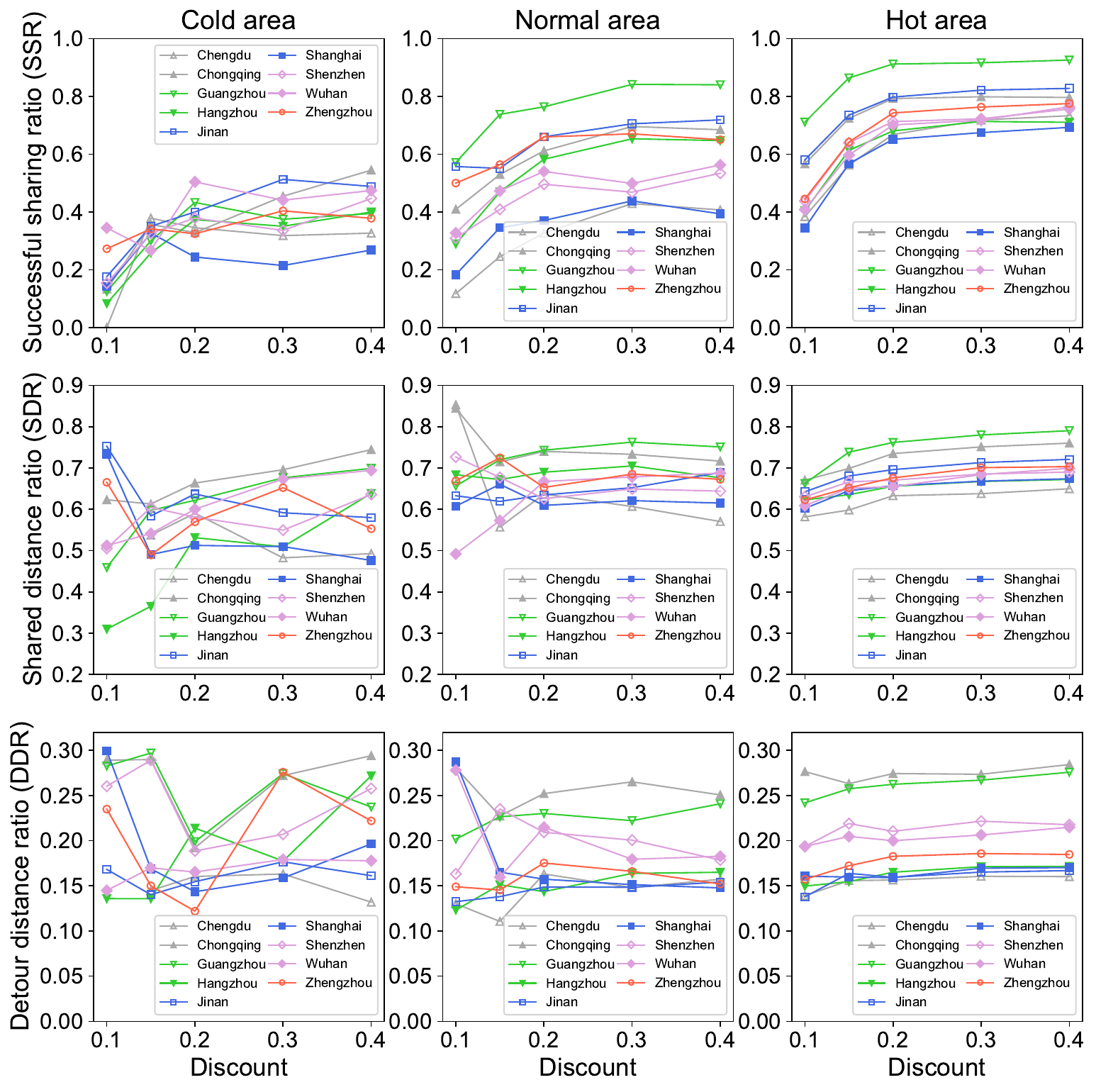}
    \caption{Spatiotemporal distribution.}
    \label{fig: ST_distribution}
\end{figure}

These results indicate that TNCs should restrict ride-sharing services in cold areas to prevent revenue loss since the SSR in cold areas is significantly low. In addition, TNCs cannot dramatically improve the SSR or SDR by increasing the discount rate for ride-sharing when it is higher than 20\%, which is consistent with the outcomes presented in Section \ref{sec: overall_results}. Also, the above results reveal the factors that can lead to a loss of revenue when introducing ride-sharing. The first factor is the low SSR, leading to a discounted charge to a few customers choosing ride-sharing but without improving the service efficiency. Another factor is the saved distance from ride-sharing is limited (the explicit results are presented in Section \ref{sec: sustainability}). On the one hand, if customers choose ride-sharing but the platform fails to pool them with others, they do not save travel distances or vehicle miles traveled (VMT) compared with solo-hailing. On the other hand, even if the platform pools customers successfully, the reduced travel distance is relatively limited on average. For example, according to the results depicted in Figure \ref{fig: ST_distribution}, we assume the SDR and DDR for a customer are 0.6 and 0.2, respectively. Remember that these ratios are calculated based on customers' original trip distances without ride-sharing. As a result, the saved distance ratio for the customer is calculated as 1 - (1+0.2-0.6/2) = 0.1, which is relatively low. Therefore, the saved distance from ride-sharing may not compensate for the corresponding discount for customers choosing ride-sharing, leading to a loss of revenue.

\subsection{Economic and environmental sustainability}\label{sec: sustainability}

A lot of existing studies have investigated charge and reward strategies for carbon emissions from different mobility services and their implications on customers' choices \citep{hong2022optimal, ding2023credit, wang2023low}. This section aims to find out whether the economic benefits of carbon emission reductions resulting from ride-sharing can compensate for its revenue loss. Figure \ref{fig: saved_CO2} (a) shows the average saved distance of each customer choosing ride-sharing with different discounts across all cities. Remind that the saved distance of customers who are willing to participate in ride-sharing but the platform fails to pool them with others is 0. Even with a 0.4 discount rate for ride-sharing, a customer can save the travel distance by no more than 800 meters on average, indicating the saved distance by introducing ride-sharing is relatively limited. Furthermore, the average reduced CO$_2$ emissions across all cities are calculated, as shown in Figure \ref{fig: saved_CO2} (b). The saved CO$_2$ emissions for each trip increase from around 40 g to 100 g as the discount for ride-sharing increases from 0.1 to 0.4, demonstrating that attracting more customers to ride-sharing can reduce more carbon emissions. However, according to the data reported by the World Bank, the average carbon price in China in 2024 is 12.6 US\$ (around 90 CNY) per ton of CO$_2$ emissions\footnote{Data source: https://carbonpricingdashboard.worldbank.org/compliance/price. Access date: December 4th, 2024.}. As a result, the average monetary benefit of carbon emission reductions brought by ride-sharing is no more than 0.01 CNY for each trip, which definitely cannot compensate for the revenue loss compared with solo-hailing.

\begin{figure}[!ht]
    \centering
    \subfigure[Saved distance in each city]{\includegraphics[width=0.4\linewidth]{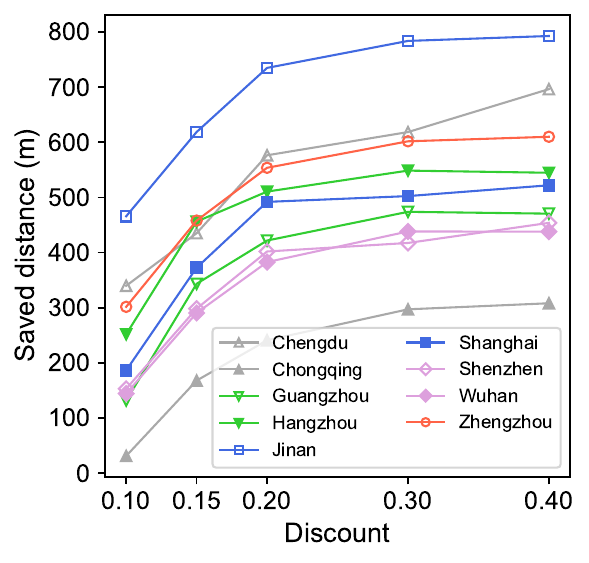}}
    \subfigure[Average saved CO$_2$ emissions]{\includegraphics[width=0.4\linewidth]{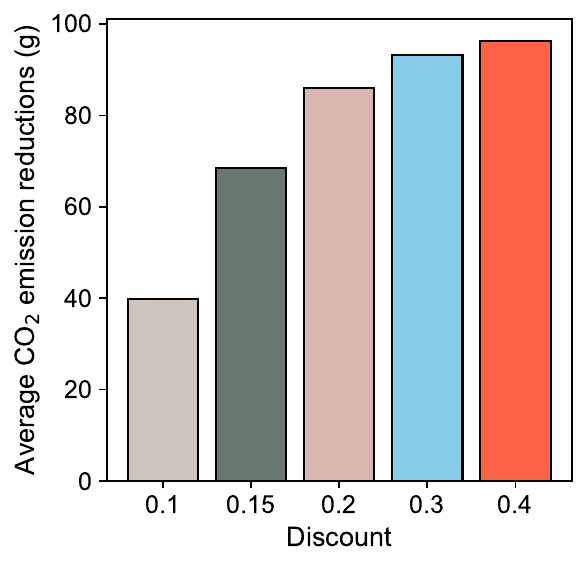}}
    \caption{Saved distance and CO$_2$ emissions resulting from ride-sharing.}
    \label{fig: saved_CO2}
\end{figure}

The above results indicate that the monetary rewards from the environmental benefits of introducing ride-sharing are insignificant, which almost does not contribute to maintaining the economic sustainability of ride-sharing services. Therefore, from the perspective of the market economy, the carbon credit mechanism may not significantly affect customers' choices between ride-sharing and solo-hailing. In this context, governments can subsidize ride-sharing services to maintain their environmental and social benefits.

\section{Discussion}\label{sec: discussion}

\subsection{Scenario analysis}

This section analyzes a typical scenario to further demonstrate the reason for the monetary loss when introducing the ride-sharing service. As shown in Figure \ref{fig: scenario_analysis}, consider an idle vehicle and two customers $C_1$ and $C_2$ with original trip distances of $L_1$ and $L_2$, respectively. Without loss of generality, we assume $L_1 \geq L_2$, denoting the original price of $C_1$ is higher than $C_2$. In a pure solo-hailing system, the platform should match the idle vehicle with $C_1$ to maximize the revenue. In this way, the revenue of unit time is calculated as follows:

\begin{equation}
    \mu^{\text{solo}} = \frac{w_1}{(x_1 + L_1) / v},
\end{equation}
where $w_1 = k_1 + k_2L_1$ denotes the trip fare without discount for $C_1$. If ride-sharing is introduced, however, the two customers can be assigned to the idle vehicle simultaneously. In this context, the platform needs to consider four different scenarios involving two customers' choices between solo-hailing and ride-sharing and estimate the expected revenue. But in this section, only the scenario where two customers choose ride-sharing is discussed for simplicity. The shortest route for the vehicle to pick up and deliver the customers is annotated by the solid arrows in Figure \ref{fig: scenario_analysis}. Therefore, the revenue from pooling the two customers is calculated as follows:

\begin{equation}
    \mu^{\text{share}} = \frac{(1-\theta)(w_1 + w_2)}{(x_1 + x_2 + x_3 + x_4) / v},
\end{equation}
where $w_1$ and $w_2$ represent the original trip fares for $C_1$ and $C_2$, respectively.  To prevent monetary loss, we should ensure that the revenue of unit time from pooling the two customers is not lower than that from pure solo-hailing, as follows:

\begin{equation}
    \frac{\mu^{\text{share}}}{ \mu^{\text{solo}}} = \frac{(1-\theta)(1 + w_2/w_1)}{(x_1 + x_2 + x_3 + x_4) / (x_1 + L_1)} \geq 1.
\end{equation}
Therefore, the revenue ratio $\mu^{\text{share}} / \mu^{\text{solo}}$ depends on the discount rate for ride-sharing $\theta$, trip fare ratio of two customers $w_2/w_1$, and trip distance ratio of the shared trip to the solo-hailing trip $(x_1 + x_2 + x_3 + x_4) / (x_1 + L_1)$.

\begin{figure}[!ht]
    \centering
   \includegraphics[width=0.5\linewidth]{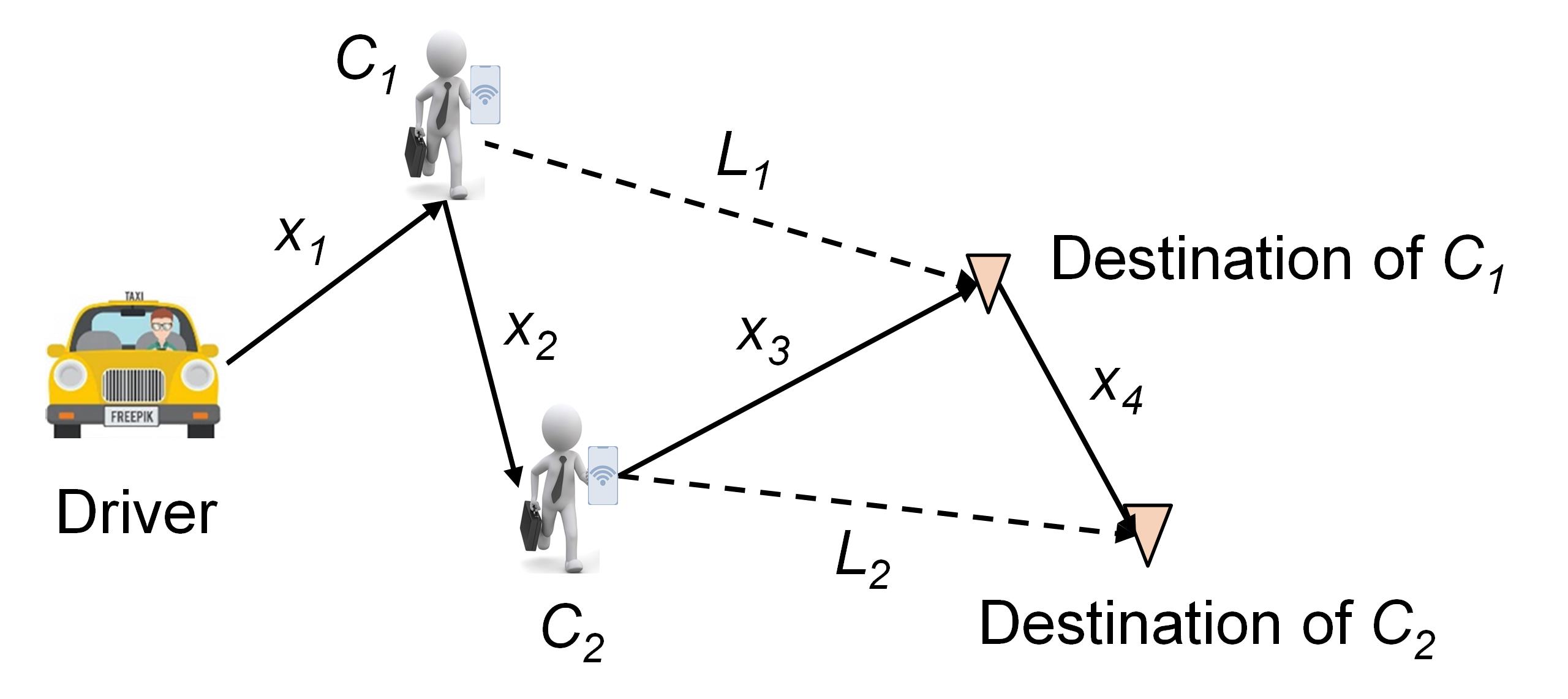}
    \caption{Illustration of scenario analysis.}
    \label{fig: scenario_analysis}
\end{figure}

Figure \ref{fig: revenue_ratio} depicts the relationship between the revenue ratio and three dependent factors. In particular, if the trip fare ratio is low, e.g., $w_2/w_1 = 0.4$, then introducing ride-sharing can induce revenue loss except for the scenario where customers are willing to choose ride-sharing with a low discount (e.g., $\theta = 0.1$) and the trip distance ratio is low (e.g., 1.2). This indicates that pooling customers with trip fares varying significantly can result in monetary loss compared to accommodating customers with a higher trip fare using the solo-hailing service. In this context, the platform should offer a low discount for ride-sharing and pool customers with similar itineraries as much as possible to prevent reducing revenue. When the trip fares for customers are identical (i.e., $w_2/w_1 = 1.0$), the platform can earn more revenue by pooling the two customers even with a relatively higher discount for ride-sharing (e.g., $\theta = 0.2$) and a longer detour distance (e.g., trip distance ratio equals 1.4). But still, if the platform offers a 30\% or even higher discount for ride-sharing, the revenue from ride-sharing is likely to be lower than that from solo-hailing. This finding is consistent with the conclusions discussed in Sections \ref{sec: overall_results} and \ref{sec: st_analysis}. In addition, a larger trip distance ratio reduces the revenue of unit time from ride-sharing, which can also represent the implications of the detour ratio on ride-sharing. This demonstrates the trade-off between the mobility efficiency and sharing probability of ride-sharing services. On the one hand, a larger detour ratio can increase the sharing probability for customers since the detour constraint is less restrictive. On the other hand, however, the longer detour distance affects the efficiency of ride-sharing, decreasing the revenue of unit time. Therefore, the guaranteed detour ratio should be designed a reasonable value in real-world applications.

\begin{figure}[!ht]
    \centering
    \includegraphics[width=0.8\linewidth]{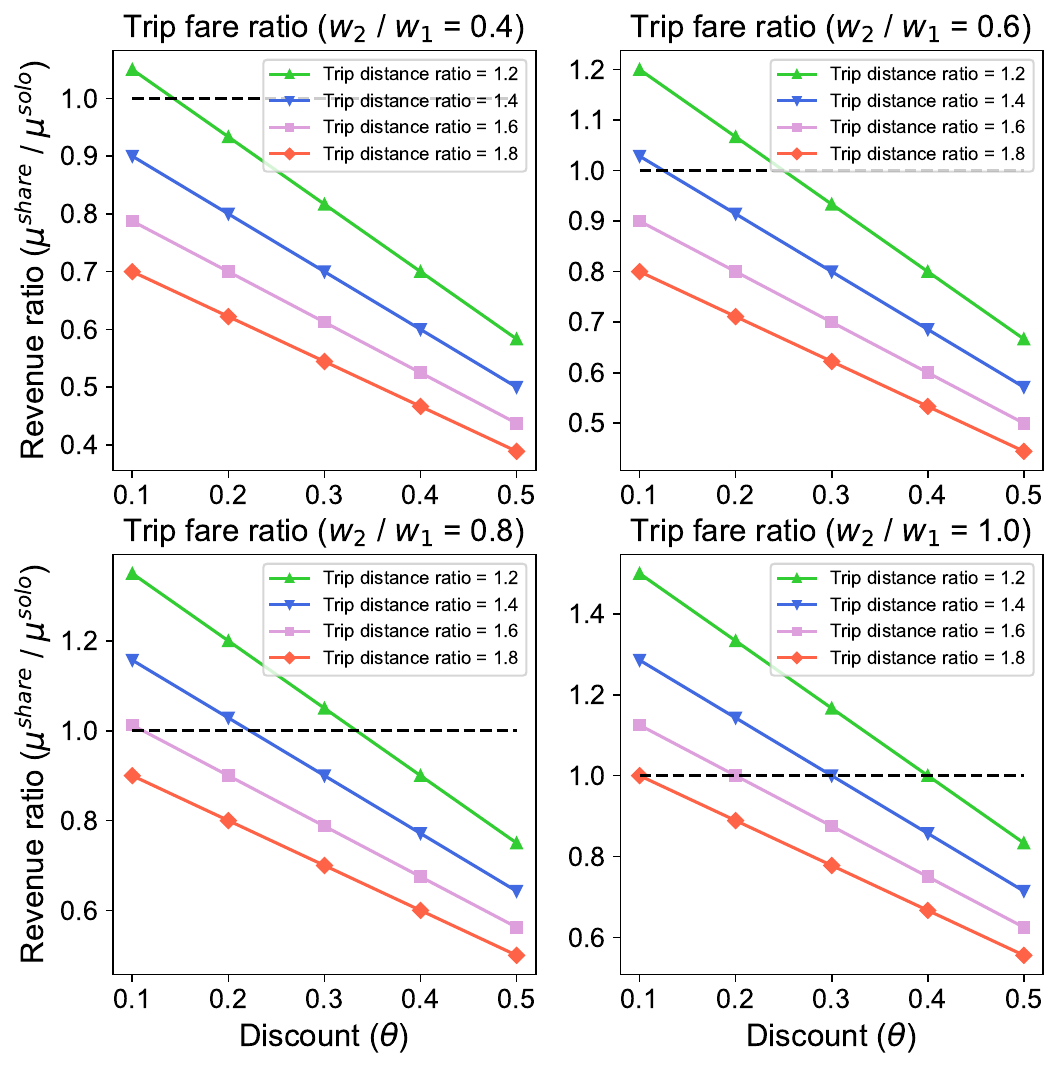}
    \caption{Relationship among the revenue ratio, trip fare ratio, discount for ride-sharing, and travel distance ratio.}
    \label{fig: revenue_ratio}
\end{figure}

The above analysis further demonstrates the reason why introducing ride-sharing can induce revenue loss. Besides the failure to pool customers can result in monetary loss, successfully pooling customers with significantly varied trip fares may also lead to less revenue. This scenario analysis is simple but representative of describing the characteristics of ride-sharing, providing a few valuable references for designing effective and efficient ride-sharing strategies. First, the platform should prioritize pooling customers with similar trip fares and itineraries to prevent revenue loss. Second, the platform should offer a relatively low discount (e.g., no more than 20\%) for ride-sharing in most cases. Last but not least, the platform should guarantee a reasonable detour ratio for customers choosing ride-sharing to balance the service efficiency and pooling probability.

\subsection{Ride-sharing as a public service}\label{sec: ride-sharing-public}

The nature of TNCs as private companies seeking profit maximization determines that TNCs do not fundamentally have the motivation to develop the ride-sharing service because of the revenue loss. However, the fare is an important factor influencing customers' adoption of the ride-sharing service, and the service performs better as the customers grow. It puts ride-sharing in a low development dilemma, even with policy encouragement. On the other hand, although introducing ride-sharing cannot increase the revenue of TNCs, it can benefit society by improving the service rate and lowering the waiting time and carbon emissions. Therefore, this study argues that ride-sharing should be incorporated as a public mobility service with suitable subsidies to solve its development dilemma.

From the perspective of occupancy, in a narrow sense, traditional taxi or solo-hailing services are seldom considered public mobility services because they serve single customers with low occupancy instead of a customer group. However, differing from the above services, ride-sharing has shown its ability to significantly improve vehicle occupancy by 16.7\% with a maximum of two customers. Meanwhile, we need to notice that ride-sharing has significant economies of scale, where the occupancy further increases as the ridership grows. On the other hand, a high occupancy ride-sharing can have greater capability to improve the occupancy, e.g. serving 98\% of customers with a 23\% fleet size \citep{alonso2017demand}. Therefore, ride-sharing has promising comparability to traditional public transport in terms of occupancy.

In terms of service performance, compared to public transport, ride-sharing provides door-to-door on-demand services and achieves better accessibility \citep{cats2022beyond}, especially for communities with low public transport coverage. For example, it provides essential mobility for minorities during COVID-19 \citep{brown2023equity}. Therefore, the trip can be more comfortable, relaxing, and faster even considering a detour.

For economic, conventional public mobility services often rely on public subsidies to operate. In particular, the bus service in lots of cities heavily relies on financial subsidies. For example, the average bus subsidy in 33 cities in China from 2016-2019 was about 261.16 million USD \citep{liu2024dynamic}, while the average public fund spent on each of the US Top 50 bus agencies in 2021 was about 360 million USD \footnote{https://www.transit.dot.gov/sites/fta.dot.gov/files/2022-10/Top\%2050\%20Profiles.pdf}. However, even with the subsidy, the declining trend in bus ridership has been prevalent in global cities in recent years, especially after COVID-19 \citep{yu2024rethinking}. Since revenue is the main factor influencing the customers’ adoption of ride-sharing, and TNCs have limited capability for discounts, we argue whether it is possible to use a portion of the funds originally allocated for maintaining public transit operations, which have been minimally effective, to subsidize ride-sharing, as the latter represents a trend for the future. For example, substituting bus routes with low occupancy that heavily rely on public funds with ride-sharing and subsidizing the trip to offer an affordable and efficient trip to customers, profitable service for TNC, and environmentally sustainable outcome to the society.

Recently, some studies discussed the possibility of integrating ride-sharing into the public transport system. For example, \cite{gurumurthy2020first} found that the incorporation of ride-sharing as a first/last mile solution helps to increase the service coverage, and walking distance, and attracts more customers. \cite{fayed2023utilization} discussed the possibility of allowing ride-sharing vehicles to use the underutilized bus lanes to improve efficiency. Therefore, this study believes incorporating ride-sharing as a public service and providing subsidies can be a suitable and promising solution to solve the development dilemma of ride-sharing development, which requires further exploration.

\section{Conclusion}\label{sec: conclusion}

Based on the real-world mobility datasets from nine cities and customers' price and detour elasticity, this study conducts massive numerical experiments involving solo-hailing and ride-sharing services. The experimental results reveal the dilemma of ride-sharing: compared with solo-hailing, introducing ride-sharing can benefit social welfare but also lead to a loss of revenue. The experiment and scenario analysis results demonstrate the three factors that can result in revenue loss. First, the successful sharing ratio for customers choosing ride-sharing is low in cold areas, reducing trip fares charged to customers but not improving service efficiency. Second, the average saved (shared) trip distance between customers from ride-sharing services is limited, which may not compensate for the discount for customers. Third, pooling two customers with significantly different trip fares can also lead to a loss of revenue compared with solo-hailing which accommodates only the customer engaged at a higher price without a discount. In addition, this study finds that the direct economic benefits of carbon emission reductions brought by ride-sharing services are not substantial and cannot significantly affect customers' choices between solo-hailing and ride-sharing. Instead, TNCs should design efficient and effective pricing strategies, e.g., dynamic pricing, to prevent the revenue loss of ride-sharing. Also, governments can subsidize ride-sharing services to maintain or even enlarge the social welfare from ride-sharing, including reducing carbon emissions and customers' waiting times, increasing vehicle occupancy rates and customers' service rates, etc.

This study paves the path for a few research directions. First, customers' elasticity of waiting time can be further calibrated, and how to efficiently and accurately estimate their waiting time during simulations can also be studied. Second, a few effective pricing algorithms can be proposed to increase ride-sharing revenue, e.g., dynamic pricing and high-capacity ride-sharing services. Last but not least, how to leverage the government subsidy to enhance ride-sharing services can be considered. 

\newpage
\bibliography{reference}

\newpage

\appendix

\section{Experimental results of other scenarios}\label{seca: exp_results}

\begin{figure}[!ht]
    \centering
    \subfigure[Revenue]{\includegraphics[width=0.32\linewidth]{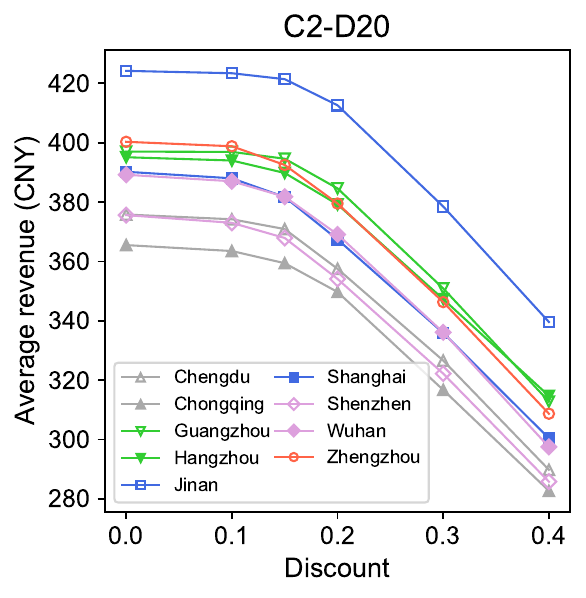}}
    \subfigure[Service rate]{\includegraphics[width=0.32\linewidth]{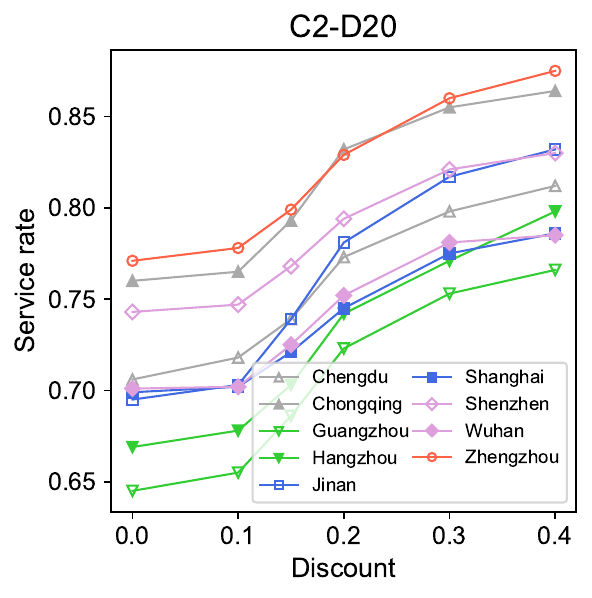}}
    \subfigure[Service rate]{\includegraphics[width=0.32\linewidth]{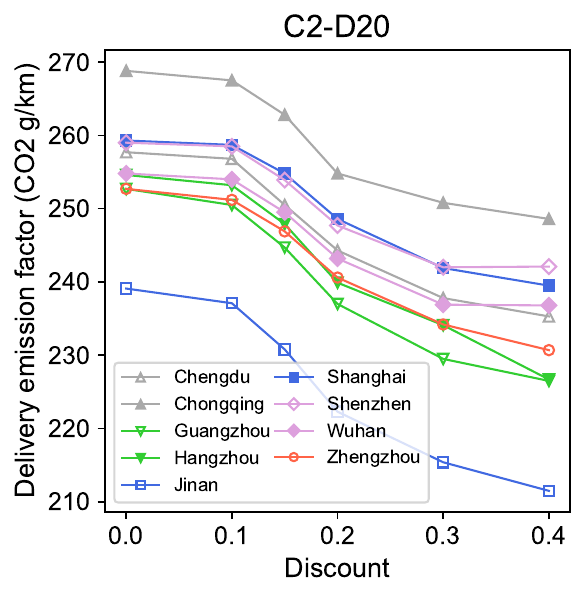}}
    \subfigure[Vehicle occupancy rate]{\includegraphics[width=0.32\linewidth]{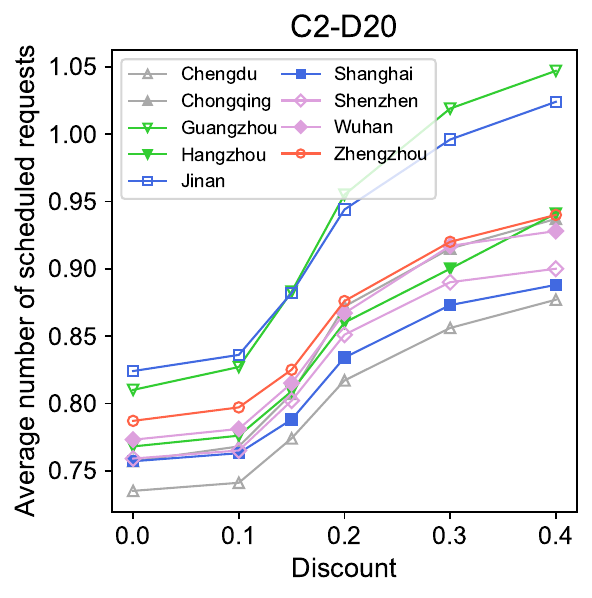}}
    \subfigure[Pickup time]{\includegraphics[width=0.32\linewidth]{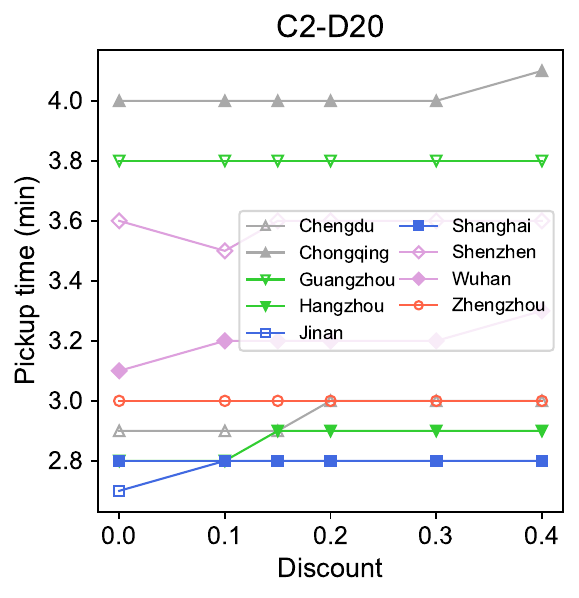}}
    \subfigure[Matching time]{\includegraphics[width=0.32\linewidth]{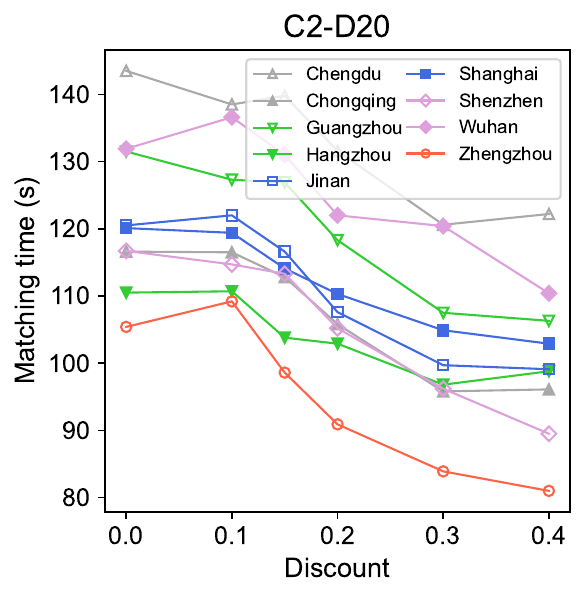}}
    \caption{Experimental results involving 20\% maximum detour ratio.}
    \label{fig: C2-D20}
\end{figure}

\begin{figure}[!ht]
    \centering
    \subfigure[Revenue]{\includegraphics[width=0.32\linewidth]{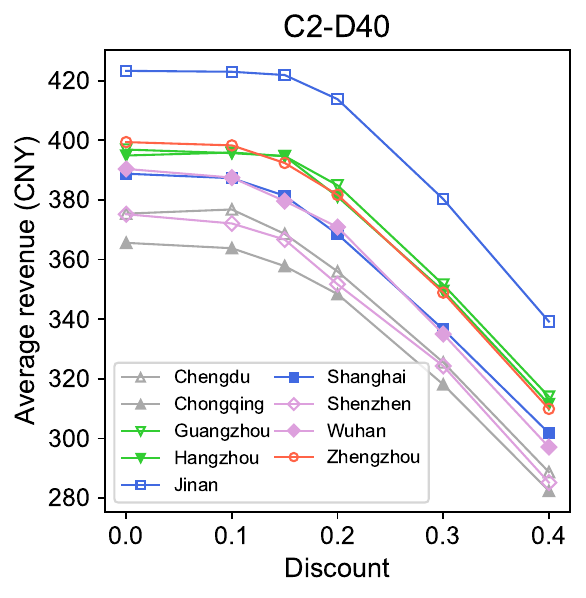}}
    \subfigure[Service rate]{\includegraphics[width=0.32\linewidth]{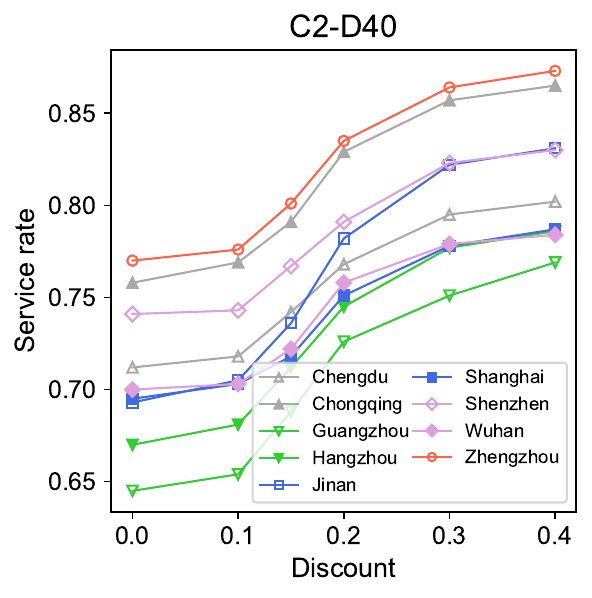}}
    \subfigure[Service rate]{\includegraphics[width=0.32\linewidth]{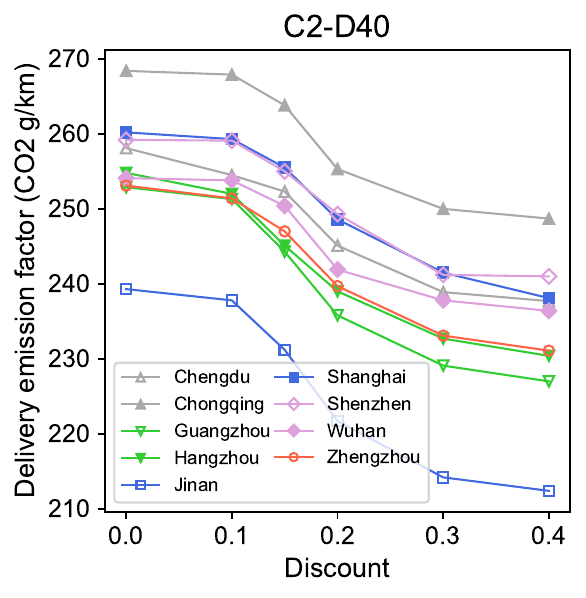}}
    \subfigure[Vehicle occupancy rate]{\includegraphics[width=0.32\linewidth]{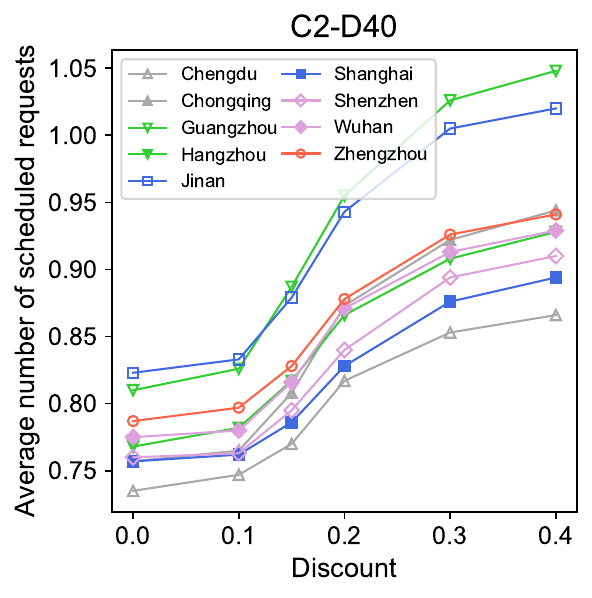}}
    \subfigure[Pickup time]{\includegraphics[width=0.32\linewidth]{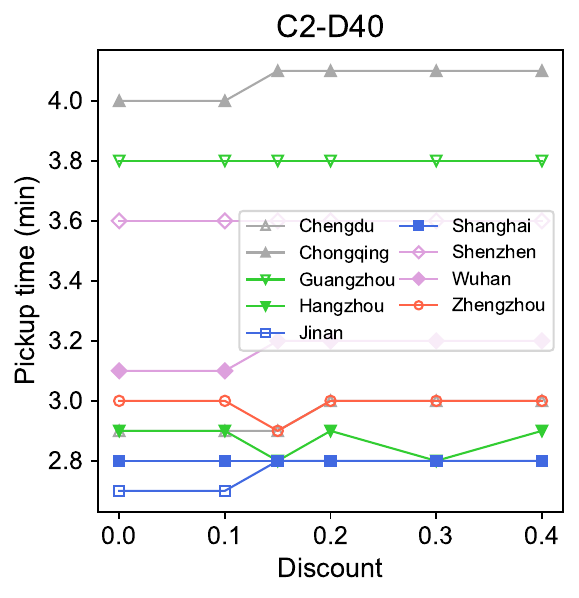}}
    \subfigure[Matching time]{\includegraphics[width=0.32\linewidth]{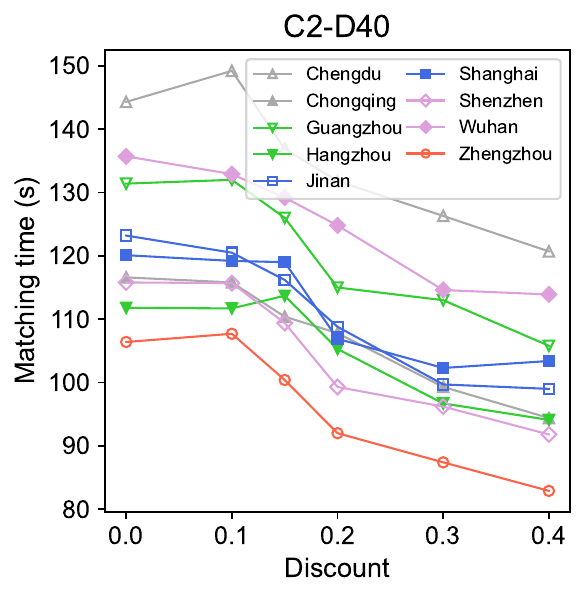}}
    \caption{Experimental results involving 40\% maximum detour ratio.}
    \label{fig: C2-D40}
\end{figure}

\end{document}